\documentclass[journal, 12pt, onecolumn, draftclsnofoot]{IEEEtran}
\usepackage{graphicx}
\usepackage{setspace}
\usepackage{amssymb}
\usepackage{amsfonts}
\usepackage{mathrsfs}
\usepackage{amssymb,amsmath}
\usepackage{epstopdf}
\usepackage{bm}
\usepackage{algorithm}
\usepackage{algorithmic}
\usepackage{amsmath,amssymb,epsfig,graphics,subfigure}
\usepackage{flushend}
\usepackage{theorem}
\usepackage{array,color}
\usepackage[compress]{cite}
\usepackage{stfloats}
\usepackage{graphicx}
\usepackage{subfigure}
\usepackage{flushend}
\usepackage{array}
\usepackage{cases}
\usepackage{multirow}
\usepackage{chngpage}
\usepackage{longtable}
\usepackage{lineno}
\usepackage{amsfonts}
\usepackage{mathrsfs}
\usepackage{amssymb,amsmath}
\usepackage{algorithm}
\usepackage{algorithmic}
\usepackage{amsmath,amssymb,epsfig,graphics,subfigure}
\usepackage{theorem}
\usepackage{array,color}
\usepackage{amssymb}

\theoremheaderfont{\normalfont\bfseries}
\begin{document}
\title{OTFS-SCMA: A Downlink NOMA Scheme for Massive Connectivity in High Mobility Channels}
\author{Haifeng Wen, Weijie Yuan,~\IEEEmembership{Member,~IEEE}, \\ Zilong Liu,~\IEEEmembership{Senior Member,~IEEE}, and Shuangyang Li,~\IEEEmembership{Member,~IEEE}
\thanks{
Haifeng Wen is with the information hub, The Hong Kong University of Science and Technology (Guangzhou), Guangzhou, China (email: hwen904@connect.hkust-gz.edu.cn).

Weijie Yuan is with the Department of Electrical and Electronic Engineering, Southern University of Science and Technology, Shenzhen 518055, China (email: yuanwj@sustech.edu.cn).

Zilong Liu is with the School of Computer Science and Electronic Engineering, University of Essex, 1NW.4.12, Colchester Campus, UK (e-mail: zilong.liu@essex.ac.uk).

Shuangyang Li is with the School of Electrical Engineering and Telecommunications, University of New South Wales, Sydney NSW 2052, Australia (e-mail: shuangyang.li@unsw.edu.au).}
}
\maketitle
\begin{abstract}
This paper studies a downlink system that combines orthogonal-time-frequency-space (OTFS) modulation and sparse code multiple access (SCMA) to support massive connectivity in high-mobility environments.
We propose a cross-domain receiver for the considered OTFS-SCMA system which efficiently carries out OTFS symbol estimation and SCMA decoding in a joint manner. 
This is done by iteratively passing the extrinsic information between the time domain and the delay-Doppler (DD) domain via the corresponding unitary transformation to ensure the principal orthogonality of errors from each domain.
We show that the proposed OTFS-SCMA detection algorithm exists at a fixed point in the state evolution when it converges.
To further enhance the error performance of the proposed OTFS-SCMA system, we investigate the cooperation between downlink users to exploit the diversity gains and develop a distributed cooperative detection (DCD) algorithm with the aid of belief consensus. 
Our numerical results demonstrate the effectiveness and convergence of the proposed algorithm and show an increased spectral efficiency compared to the conventional OTFS transmission.

\end{abstract}


\section{Introduction}
\subsection{Background}
As the 5G networks commercially rolling out across the world, the study of the mobile communication systems in the beyond 5G (B5G) era has received much attention \cite{Saad2020}. 
In the future wireless networks, the proliferation of connected autonomous vehicles and  unmanned aerial vehicles, as well as the integration of terrestrial and non-terrestrial networks (e.g., inter-satellute and satellite-to-ground communications), raise the demands for extremely reliable and rapid data services in high-mobility environments. 
While the 5G networks aim for robust communications at moving speeds up to 500 km/h, this target has been raised to 1000 km/h and higher in the current 6G research \cite{Giordani2020, Zilong2022}. The widely adopted orthogonal frequency division multiplexing (OFDM) modulation may be incapable as the system orthogonality could be severely destroyed by the increased inter-carrier interference caused by the Doppler effect \cite{Raviteja2019}.

Recently, orthogonal time frequency space (OTFS) modulation has emerged as a promising technique for robust data transmission over high-mobility wireless channels \cite{Hadani2017, Wei2021, Li2022CL, Yuan2022CL}. Compared to the OFDM modulation adopting the time-frequency (TF) domain symbol multiplexing, OTFS modulation considers the signal representation in the delay-Doppler (DD) domain, in which the channel responses are relatively sparse and compact \cite{Hadani2017}. 
Instead of the conventional doubly selective channels, OTFS permits a separable and quasi-static channel in the DD domain, which can be leveraged for more efficient communication system designs \cite{Hadani2017}. 
{Besides, OTFS modulation spreads each information symbol modulated in the DD domain to the whole TF domain, thus premitting the exploitation of the full channel diversity to enhance error performance \cite{Raviteja2019Performance, Li2021Performance}.}

In addition to the requirement of high reliability, with the increasingly congested spectrum yet more stringent quality-of-service requirements, it is challenging to concurrently support a massive number of communication links in high-mobility channels.
Against this background, non-orthogonal multiple access (NOMA) has received tremendous research attention in the past years as an enabling wireless paradigm to meet the heterogeneous demands on spectral efficiency, latency, and connectivity \cite{Ding2017}. The existing dominant NOMA schemes can be divided into two major categories: power-domain NOMA and code-domain NOMA. 
Power-domain NOMA distinguishes users by assigning them with different power levels, in contrast to code-domain NOMA which relies on carefully designed codebooks/sequences. 
In particular, sparse code multiple access (SCMA) is a disruptive code-domain NOMA technique by employing different  sparse codebooks \cite{Nikopour2013,Liu2021}. An efficient message passing algorithm (MPA) can be carried out to exploit the codebook sparsity for near-optimum multiuser detection performance \cite{Hoshyar2008}. 

In this paper, we aim to address the aforementioned technical challenges by integrating OTFS and SCMA to harness the benefits of both schemes.
Besides, observed by the fact that each user also receives the information of other users in a downlink system, we further consider the use of cooperative detection. By doing so, user diversity gains can be achieved by cooperatively exchanging the detection results with different users. 
Specifically, in a cooperative network, each user only needs to share their local information with neighboring users. Such a distributed cooperative scheme has been employed in both sensing and communication systems \cite{Wymeersch2009, Meyer2016, Ng2008, Ding2015,Yuan2018}.

\subsection{Related Works}
The design of OTFS aided multiple access in a high-mobility environment has attracted much research attention recently.
An orthogonal multiple access (OMA) scheme for uplink OTFS systems was proposed in \cite{Khammammetti2019}, where the information symbols of different users are placed in a non-overlapping manner in the TF domain. 
The authors in \cite{Chong2022} demonstrated that the achievable rates of two DD domain multiple access schemes (delay-division and Doppler-division) for uplink OTFS systems are noticeably improved compared to the conventional orthogonal frequency-division multiple access (OFDMA) systems.
{By integrating power-domain NOMA with OTFS in the time domain, it was shown in \cite{Chatterjee2021} that power-domain NOMA-OTFS provides a higher sum spectral efficiency compared to OTFS-OMA. }
However, rapid real-time power allocation in high mobility channels may be impractical.
By contrast, code-domain OTFS-NOMA is attractive as the codebook/sequence assignment is independent of the specific locations of users.
{
An OTFS-SCMA scheme was developed in \cite{Deka2021} with a two-stage detector and a single-stage detector for both downlink and uplink systems, respectively. 
However, the non-iterative two-stage downlink detector that first conducts linear minimum mean square error (LMMSE) equalization and then performs MPA decoding in the DD domain may not perform well. This is because the LMMSE estimator can hardly achieve Bayes optimality with the superimposed SCMA codebook, which potentially leads to performance degradation in the subsequent MPA decoding part.
To facilitate the OTFS-SCMA detection, classical OTFS channel estimation methods can be directly applied in the downlink OTFS-SCMA systems, e.g., the threshold-based pilot-embedded method \cite{Raviteja2019Pilot}, the sparse Bayesian learning-based method \cite{Wei2022OffGrid}, the orthogonal matching pursuit (OMP)-based method \cite{Shen2019OMP}.
As for OTFS-SCMA systems, there have been two methods discussed in the literature \cite{Deka2021, Thomas2022}. The authors of \cite{Deka2021} proposed a channel estimation scheme where a guard band between the pilot and the SCMA codewords was adopted to avoid interference. 
In  \cite{Thomas2022} a convolutional sparse coding-based channel estimation method was proposed by exploiting the convolution nature of OTFS input-output relation and the sparsity nature of the DD domain effective channel, where the spectral efficiency is comparable with the single user case.
}

When rectangular pulse shaping waveform is adopted, the fractional Doppler effect results in more channel coefficients in the Doppler domain, leading to dense channel response in the DD domain which in turn raises challenges for channel estimation and signal detection.
So far, very few works are known on the tackling of the aforementioned fractional-Doppler shift problem. 
{For example, a zero-padded OTFS system with Rake receivers was proposed in \cite{Thaj2020} by carrying out equalization in the delay-time domain, where the channel sparsity is not affected by the fractional Doppler shifts.}
The authors in \cite{Ge2021} considered fractionally spaced sampling in the time-domain for enhanced pulse-shaped OTFS systems.
Instead of performing detection in a single domain, a cross-domain OTFS detector was proposed in \cite{Li2021}. The idea of \cite{Li2021} is to carry out equalization in the time domain to exploit the natural sparsity of time-domain effective channels in OTFS systems, whilst performing the denoising in the DD domain. Similar to the core idea of orthogonal approximate message passing (OAMP) \cite{Ma2017,LeiLiu2021}, the cross-domain detector passes extrinsic information (e.g. means and covariance matrices) between the two domains via corresponding unitary transformation to ensure the principal orthogonality of errors from each domain. 


\subsection{Motivations and Contributions}
Despite the above-mentioned works on OTFS-aided multiple access systems \cite{Khammammetti2019, Chong2022, Chatterjee2021, Deka2021}, the enabling of massive connectivity in high mobility channels is a largely open research topic. Inspired by \cite{Li2021, Ma2017,LeiLiu2021}, we propose to perform OTFS symbol estimation and SCMA decoding in different domains to fully exploit the advantages of OTFS and SCMA.
{Furthermore, we observe that the OTFS frames of all SCMA users carry the same information symbols, and different users may experience different physical channels. Based on this observation, we conceive distributed cooperative detection (DCD)\footnote{{DCD aims to reach consensus on the global message and offers user diversity gain from all downlink users by using belief consensus \cite{Xiao2005}. Belief consensus is a belief propagation algorithm that allows distributed computation of the products of several local functions in a factor graph over the same variable node \cite{Yuan2018}.}} \cite{Yuan2018} in our proposed cross-domain OTFS-SCMA detector with a novel multi-layer detection structure to achieve user diversity gains.}
    


The main contributions of this paper are as follows:
\begin{itemize}
    \item[1)] We propose a single-layer joint OTFS-SCMA detector based on the cross-domain detection, where the OTFS equalization is conducted in the time domain with a linear minimum mean squared error (L-MMSE) equalizer and the SCMA signals are decoded in the DD domain by a conventional MPA decoder \cite{Hoshyar2008}. 
    The extrinsic information from each domain is iteratively passed and updated through unitary transformation, namely cross-domain message passing \cite{Li2021}. {With the aid of extrinsic message passing, the estimation/decoding errors in one domain are principally orthogonal to that in the other.}
    {Due to the sparsity of SCMA codebooks, the covariance matrices of decoded codewords are rank-deficient, posing difficulties for low-complexity calculation. Such difficulties may not be addressed by straightforward application of the techniques in \cite{Li2021}.}
    To proceed, we assume that the superimposed codewords in each DD domain resource are independent and identically distributed (i.i.d.). The covariance matrices can thus be regarded as diagonal, yielding a simple calculation of the matrix inverse.
    \item[2)] We extend the single-layer OTFS-SCMA detector to a novel multi-layer one by allowing cooperation between downlink users, which is performed over all downlink users while the single-layer detection is performed in one downlink user only.
    Specifically, we first propose a \textit{joint} cross-domain and DCD, where iterative belief consensus between each layer (user) is conducted in each cross-domain message passing iteration (after the time domain equalization). 
    Different from the \textit{joint} structure, we consider a \textit{separate} structure in which we perform several iterations of belief consensus, followed by an additional MPA decoding after the local OTFS-SCMA cross-domain detection. 
    Furthermore, to further reduce the energy consumption during cooperation, we present a reduced belief consensus scheme that only broadcasts partial local information. 
    \item[3)] {We analyze the state evolution (SE) \cite{Ma2015} of the proposed OTFS-SCMA cross-domain detector and derive a fixed point of SE. We prove that when the detector converges, the average mean square error (MSE) values of the time domain coincide with that in the DD domain at the fixed point. The existence of the fixed point means that the proposed detector potentially achieves Bayes optimality, i.e., the proposed detector can converge to the MMSE if there is exactly one fixed point for SE.}
    
\end{itemize}


\subsection{Notations}
$\mathbb{C}^{k\times n}$ denotes the $(k\times n)$-dimensional complex matrix spaces;
$\mathbf{X}^\text{T}$ and $\mathbf{X}^\text{H}$ denote the transpose and the Hermitian transpose of matrix $\mathbf{X}$;
$\otimes$ denotes the Kronecker product;
$\propto$ denotes equality up to a constant normalization factor;
$\text{tr}(\mathbf{X})$ and $\text{diag}(\mathbf{X})$ give the trace and a vector composed of the main diagonal elements of the matrix $\mathbf{X}$. $\mathbf{I}_{M}$ is an identity matrix with size $M\times M$;
$\mathbf{X}[i,j]$ denotes the entry in row $i$ and column $j$ of the matrix $\mathbf{X}$;
$|\cdot|$ denotes the modulus of a complex number or the cardinality of a set;
$\mathbb{E}[\cdot]$ denotes the expectation operator;
$\text{vec}(\cdot)$ denotes the vectorization operator;
$\delta(\cdot)$ represents the Dirac delta function;
$(\cdot)^{a,\text{T}}, \  (\cdot)^{p,\text{T}}, \  (\cdot)^{e,\text{T}}$ denote the \textit{a priori}, the \textit{a posterior}, and the extrinsic information for time domain, respectively;
The counterparts of DD domain are denoted as $(\cdot)^{a,\text{DD}}, \  (\cdot)^{p,\text{DD}}, \  (\cdot)^{e,\text{DD}}$; The mean and the covariance matrix of variable $\mathbf{x}$ are denoted as  $\mathbf{m}_{\mathbf{x}}$ and $\mathbf{C}_{\mathbf{x}}$, respectively.

\section{Preliminaries}

\begin{figure}[t]
    \begin{minipage}[t]{0.48\linewidth}
        \centering
        \includegraphics[width=3in]{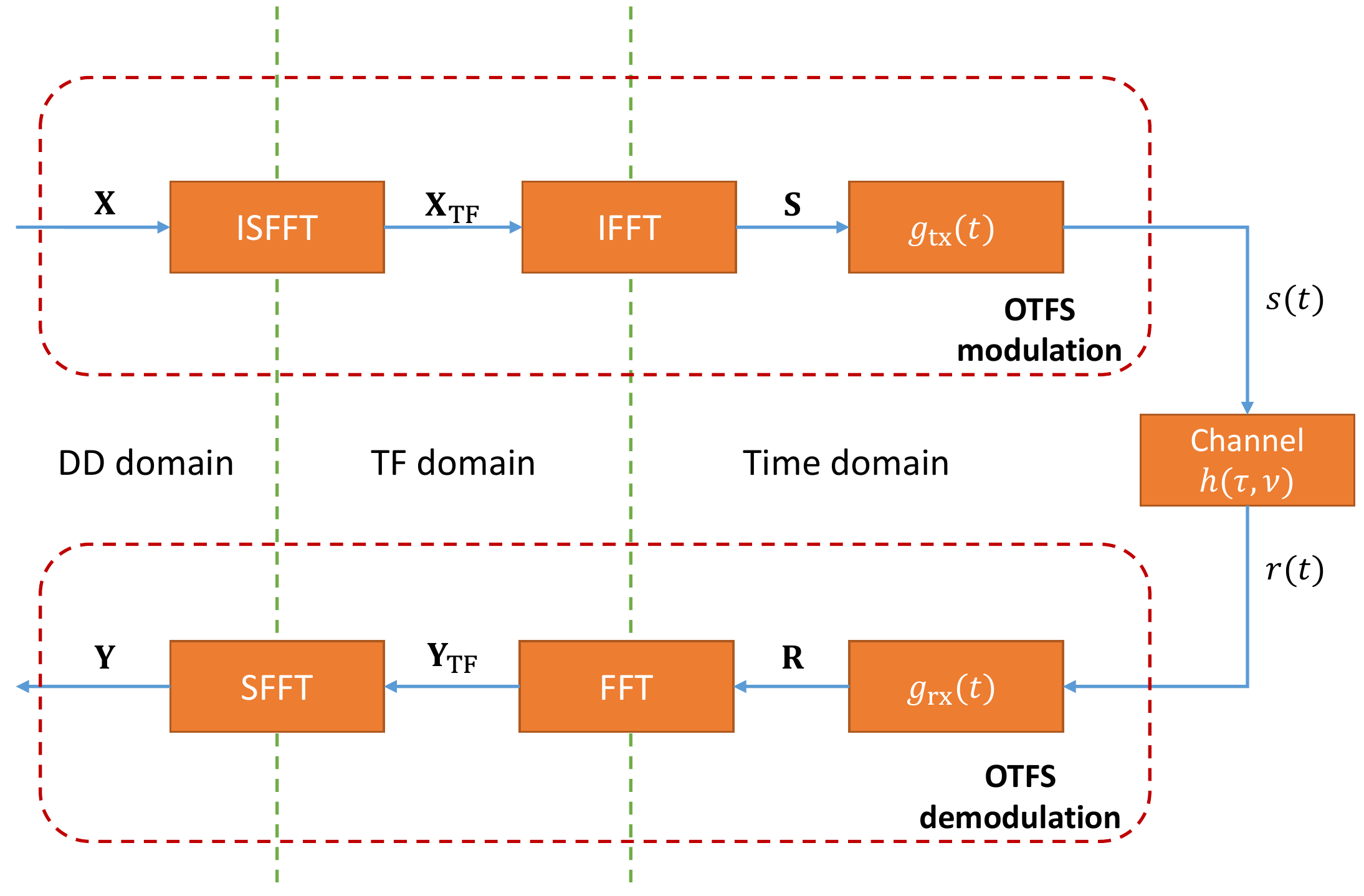}\\
        \caption{A system diagram of OTFS modulation/demodulation.}
        \label{OTFS system}
    \end{minipage} \ 
    \begin{minipage}[t]{0.48\linewidth}
        \centering
        \includegraphics[width=3in]{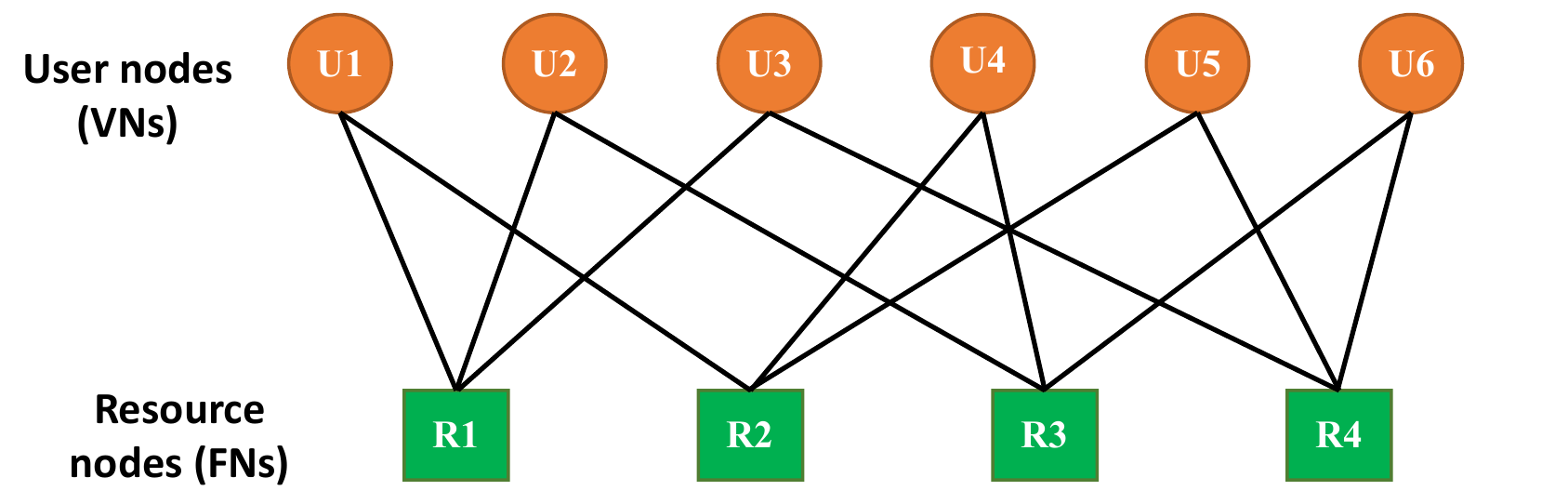}\\
        \caption{Factor graph for an $J=6$, $K=4$ SCMA system with $d_v=2$, $d_c=3$.}
        \label{Factor graph}
    \end{minipage}
    \vspace{-5mm}
\end{figure}

{
 
\subsection{OTFS}
We consider an OTFS system in Fig. \ref{OTFS system} that transmits symbols $X[m, n]$ over the DD domain grid $\Gamma = \Big\{ \left(\frac{m}{M\Delta f}, \frac{n}{NT}\right),$ $m = 0,...,M-1,$ $n = 0,...,N-1 \Big\}$, where $M\Delta f$ is the bandwidth of an OTFS frame, and $NT$ is the OTFS frame duration with $\Delta f = 1/T$. The OTFS transmitter first maps the symbols $X[m,n]$ to the TF domain grid $\Pi = \{ \left(l\Delta f, kT\right),$ $l = 0,...,M-1,$ $ k = 0,...,N-1\}$ via inverse finite symplectic Fourier transform (ISFFT) as \cite{Hadani2017} as follows:
\begin{equation}
    X_{\text{TF}}[l, k]=\frac{1}{\sqrt{NM}}\sum_{n=0}^{N-1}\sum_{m=0}^{M-1}{X[m,n]e^{j2\pi\left(\frac{nk}{N}-\frac{ml}{M}\right)}},
    \label{DD2TF}
\end{equation}
where $X_{\text{TF}}[l, k]$ denotes The TF domain transmitted symbols. The time-domain signal $s(t)$ can thus be produced by the conventional OFDM modulator, i.e., the TF domain symbols $X_{\text{TF}}[l,k]$ are then converted to a continuous time waveform $s(t)$ by the conventional OFDM modulator with the transmitter shaping pulse $g_{\text{tx}}(t)$ with duration $T$, i.e.
\begin{equation}
s(t)=\sum_{k=0}^{N-1}\sum_{l=0}^{M-1}{X_{\text{TF}}[l,k]g_{\text{tx}}(t-kT)e^{j2\pi l\Delta f(t-kT)}}.
\end{equation}
\vspace{-2mm}
The signal $s(t)$ is transmitted over a time-varying wireless channel characterized by the impulse response $h(\tau,\nu)$ with delay $\tau$ and Doppler $\nu$, given by
\begin{equation}
h(\tau,\nu)=\sum_{i=1}^P{h_i\delta(\tau-\tau_i)\delta(\nu-\nu_i)},
\label{h_ideal}
\end{equation}
where $P$ denotes the number of paths, $h_i$, $\tau_i$, and $\nu_i$ denote the path gain, delay, and Doppler shift of the $i$-th path, respectively. Note that $\tau_i$ and $\nu_i$ depend on the delay and Doppler indices of the $i$-th path, which are given by \cite{OTFSBook2022}
\begin{equation}
\tau_i=\frac{l_i}{M\Delta f}, \quad \nu_i=\frac{k_i+\kappa_i}{NT}
\end{equation}
with the integers $l_i$, $k_i$, and  the fractional Doppler term $-1/2\le\kappa_i \le 1/2$.
At the receiver side, the received signal $r(t)$  is given by
\begin{equation}
\begin{aligned}
r(t) &= \int\int{h(\tau,\nu)s(t-\tau)e^{j2\pi\nu(t-\tau)}} \,d\tau \, d\nu +n(t) \\
&= \sum_{i=1}^P{h_i s(t-\tau_i)e^{j2\pi\nu_i(t-\tau_i)}} + n(t),
\end{aligned}
\label{r(t)}
\end{equation}
where $n(t)$ is the additive white Gaussian noise (AWGN) signal with one-sided power spectral density $N_0$. After receiving $r(t)$, the OTFS receiver converts the time domain signal to the TF domain symbols $Y_{\text{TF}}[l,k]$ with the  matched filter $g_{\text{rx}}(t)$, i.e. \cite{Hadani2017}
\begin{equation}
Y_{\text{TF}}[l,k] = \int{ r(t)g^{*}_{\text{rx}}(t-kT)e^{-j2\pi l \Delta f (t-kT)} \, dt}.
\end{equation}
Finally, the DD domain received symbols $Y[m, n]$ can be obtained by performing the SFFT to $Y_{\text{TF}}[l,k]$ as
\begin{equation}
Y[m, n] = \frac{1}{\sqrt{NM}} \sum_{k=0}^{N-1} \sum_{l=0}^{M-1} Y_{\text{TF}}[l,k] e^{-j 2 \pi \left(\frac{nk}{N} - \frac{ml}{M} \right)} + \tilde{n}[m,n],
\end{equation}
where $\tilde{n}[m,n]$ represents the corresponding AWGN sample in the DD domain. 

Next, we can rewrite the input-output relationship of different domains into a vectorized form to simplify the subsequent derivation. In this paper, we follow the same notations for the matrix/vector representation of the OTFS system in \cite{Thaj2020}. Let $\mathbf{X}$, $\mathbf{Y} \in \mathbb{C}^{M\times N}$ be the transmitted and received DD domain symbol matrices and the counterparts of TF domain are denoted by $\mathbf{X}_{\text{TF}}\in\mathbb{C}^{M\times N}$ and $\mathbf{Y}_{\text{TF}}\in\mathbb{C}^{M\times N}$, respectively. For the time domain, the transmitted symbol matrix and received symbol matrix are denoted by $\mathbf{S}\in\mathbb{C}^{M\times N}$ and  $\mathbf{R}\in\mathbb{C}^{M\times N}$, respectively.

Let $\mathbf{F}_M$ and $\mathbf{F}_N$ be the normalized $M$-point and $N$-point discrete Fourier transform (DFT) matrices. Eq. (\ref{DD2TF}) can be reformulated as
\begin{equation}
\mathbf{X}_{\text{TF}} = \mathbf{F}_M \mathbf{X} \mathbf{F}_N^{\text{H}}.
\label{DD2TFm}
\end{equation}
The transmitted time domain signal from the TF domain samples with the rectangular pulse can be rewritten as
\begin{equation}
\mathbf{S}=\mathbf{I}_M \mathbf{F}_M^{\text{H}} \mathbf{X}_{\text{TF}} = \mathbf{X}\mathbf{F}_N^{\text{H}}.
\label{TF2Tm}
\end{equation}
Thus, the time domain transmitted vector $\mathbf{s}\in \mathbb{C}^{NM\times 1}$ is given by
\begin{equation}
\mathbf{s}\overset{\underset{\Delta}{}}{=}\text{vec}\left(\mathbf{S}\right)=(\mathbf{F}_N^{\text{H}} \otimes \mathbf{I}_M)\mathbf{x},
\label{DD2Tv}
\end{equation}
with the notation of that $\mathbf{x} \overset{\underset{\Delta}{}}{=}\text{vec}(\mathbf{X})$.

Considering the reduced cyclic prefix frame format, at the receiver side, after discarding the CP, we can rewrite (\ref{r(t)}) in the vectorized form by discretizing the time-domain received signal at a rate $f_s=M \Delta f$ as \cite{OTFSBook2022}
\begin{equation}
r[q] = \sum_{i=1}^P{h_i e^{j2\pi \frac{(k_i+\kappa_i)(q-l_i)}{MN}}s\left[ [q-l_i]_{MN}\right]} + n[q],
\label{r[q]}
\end{equation}
where $[\cdot]_{MN}$ denotes mod-$MN$ operation and $q = 0,...,MN-1$. Therefore, the discrete time-domain input-output relation in a vector form can be given by
\begin{equation}
\mathbf{r} = \mathbf{H}_{\text{T}}\mathbf{s} + \mathbf{n},
\label{r_Hs}
\end{equation}
where the effective time-domain channel matrix $\mathbf{H}_{\text{T}}$ is given by
\begin{equation}
	\mathbf{H}_{\text{T}} = \sum_{i=1}^{P}{h_i e^{-j2\pi \frac{(k_i+\kappa_i)l_i}{MN}}\bm{\Delta}^{k_i+\kappa_i}\bm{\Pi}^{l_i}},
	\label{H_T}
\end{equation}
where $\bm{\Delta}=\text{diag}\left(\left[1, e^{j2\pi\frac{1}{MN}},...,e^{j2\pi\frac{MN-1}{MN}}\right]\right)$ is the phase rotating matrix and $\bm{\Pi}$ is the permutation matrix (forward cyclic shift) given by
\begin{equation}
	\bm{\Pi} = \begin{bmatrix}
		0&\cdots&0&1 \\ 
		1&\cdots&0&0 \\
		\vdots&\ddots&\vdots&\vdots \\
		0&\cdots&1&0
	\end{bmatrix}_{MN\times MN}.
\end{equation}
According to (6)-(10), and by assuming rectangular pulse shaping waveform, the DD domain received symbol vector $\mathbf{y} \overset{\underset{\Delta}{}}{=} \text{vec}(\mathbf{Y})$ is given by
\begin{equation}
\mathbf{y} = \left( \mathbf{F}_N \otimes \mathbf{I}_{M} \right)\mathbf{r} = \mathbf{H}_{\text{DD}}\mathbf{x} + \tilde{\mathbf{n}},
\end{equation}
where the DD-domain effective channel matrix $\mathbf{H}_{\text{DD}}$ is given by
\begin{equation}
\mathbf{H}_{\text{DD}} = \left( \mathbf{F}_N \otimes \mathbf{I}_M \right) \mathbf{H}_{\text{T}}  \left( \mathbf{F}_N^{\text{H}} \otimes \mathbf{I}_M \right).
\label{H_DD}
\end{equation}
It can be seen that the effective channel matrix $\mathbf{H}_{\text{DD}}$ in the DD domain in (\ref{H_DD}) may become dense in the presence of the fractional Doppler shifts, whereas the time domain effective channel matrix $\mathbf{H}_{\text{T}}$ in (\ref{H_T}) remains sparse. In particular, there are at most $l_{\text{max}}$ non-zero entries in each row and column of $\mathbf{H}_{\text{T}}$. The sparsity of $\mathbf{H}_{\text{T}}$ motivates us to perform equalization in the time domain. Also, we are able to perform low-complexity near maximum likelihood (ML) detection in the DD domain once the channel is equalized in the time domain.
}

\subsection{SCMA}


Consider a downlink $K\times J$ SCMA system, where $J$ users communicate over $K$ resource nodes for multiple access. Typically, $J>K$ indicates that the number of users that concurrently communicate is larger than the total number of orthogonal resources, and the overloading factor is defined by $\lambda=J/K > 1$. 

Each user is pre-assigned a codebook $\mathcal{X}_j \in \mathbb{C}^{K\times M_{\text{mod}}}$, and $j\in \{1,2,...,J\}$, consisting of $M_{\text{mod}}$ codewords with dimension of $K$. We consider the power budgets such that $\text{Tr}(\mathcal{X}_j\mathcal{X}_j^{\text{H}})/M_{\text{mod}}=1$. The SCMA encoder of user $j$ selects a codeword from codebook $\mathcal{X}_j$ corresponding to the input binary message $b_j$ with $\log_2(M_{\text{mod}})$ bits. Let the codeword for user $j$ be $X_j=[X_{j,1},X_{j,2},...,X_{j,K}]^{\text{T}} \in \mathbb{C}^{K\times 1}$. Then the codewords of $J$ users are superimposed on $K$ resource nodes, i.e.,
\begin{equation}
    \mathbf{x}_{\text{SCMA}}=\sum_{j=1}^JX_j,
\end{equation}
where $\mathbf{x}_{\text{SCMA}}\in \mathbb{C}^{K\times 1}$ is called superimposed codewords and can be regarded as points in a large superimposed constellation whose size is $M_{\text{mod}}^J$. For SCMA, codebooks are sparse, i.e. each codeword in $\mathcal{X}_j,\ \forall j$, consists of $K-d_v$ zeros and $d_v$ non-zero elements, and each resource node carries $d_c$ users' symbols.
The relationship between user nodes (VNs) and resource nodes (FNs) can be represented by a factor graph, as shown in Fig. \ref{Factor graph} with $J=6$, $K=4$, $d_v=2$, and $d_c=3$. An edge is assigned between the two nodes if and only if the $j$-th user node occupies the $k$-th resource node, i.e., $X_{j,k}\ne 0$.

An alternative representation of the factor graph is an indicator matrix, where each row indicates a specific resource node and all the non-zero entries in that row correspond to the active users on that resource node. In specific, the indicator matrix for the factor graph shown in Fig. \ref{Factor graph} is given by
\begin{equation}
    \mathbf{F}^{\text{ind}}=\begin{bmatrix}
        1 & 1 & 1 & 0 & 0& 0 \\ 
        1 & 0 & 0 & 1 & 1 & 0 \\ 
        0 & 1 & 0 & 1 & 0 & 1 \\ 
        0 & 0 & 1 & 0 & 1 & 1 \\ 
        \end{bmatrix}.
        \label{factor graph matrix}
\end{equation}

Given a channel matrix $\mathbf{H}_{\text{DD}}$, the received signal $\mathbf{y}_{\text{SCMA}}$ of a downlink SCMA system can be denoted by
\begin{equation}
    \mathbf{y}_{\text{SCMA}} 
    =\mathbf{H}_{\text{DD}}\mathbf{x}_{\text{SCMA}} + \tilde{\mathbf{n}} 
    =\mathbf{H}_{\text{DD}}\sum_{j=1}^JX_j + \tilde{\mathbf{n}},
\end{equation}
where $\tilde{\mathbf{n}}$ is the corresponding AWGN sample vector in the DD domain.

Based on the SCMA factor graph, MPA decoder can be employed to decode the SCMA codewords \cite{Nikopour2013}.
\vspace{-0.2cm}

\subsection{Allocation of SCMA codewords in the OTFS grid}


\begin{figure}[t]
    \begin{minipage}[t]{0.7\linewidth}
        \centering
        \includegraphics[width=4.5in]{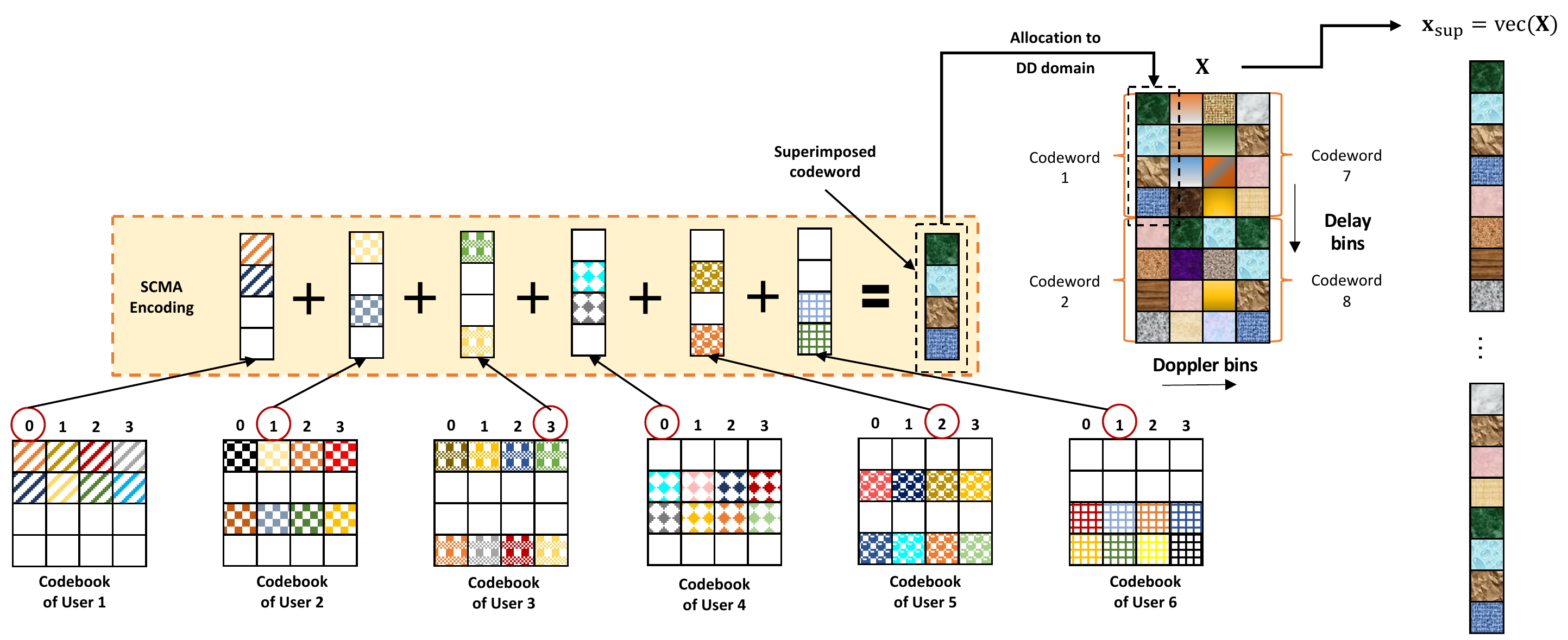}\\
        \caption{An illustration of SCMA encoding and allocation, with $M=8,\ N=4$, $M_{\text{mod}}=4$, $J=6$ and $K=4$. The codeword 1 carries the information ``013021'' of $6$ users.}
        \label{SCMA encoding and allocation}
    \end{minipage} \ 
    \begin{minipage}[t]{0.28\linewidth}
        \centering
        \includegraphics[width=1.8in]{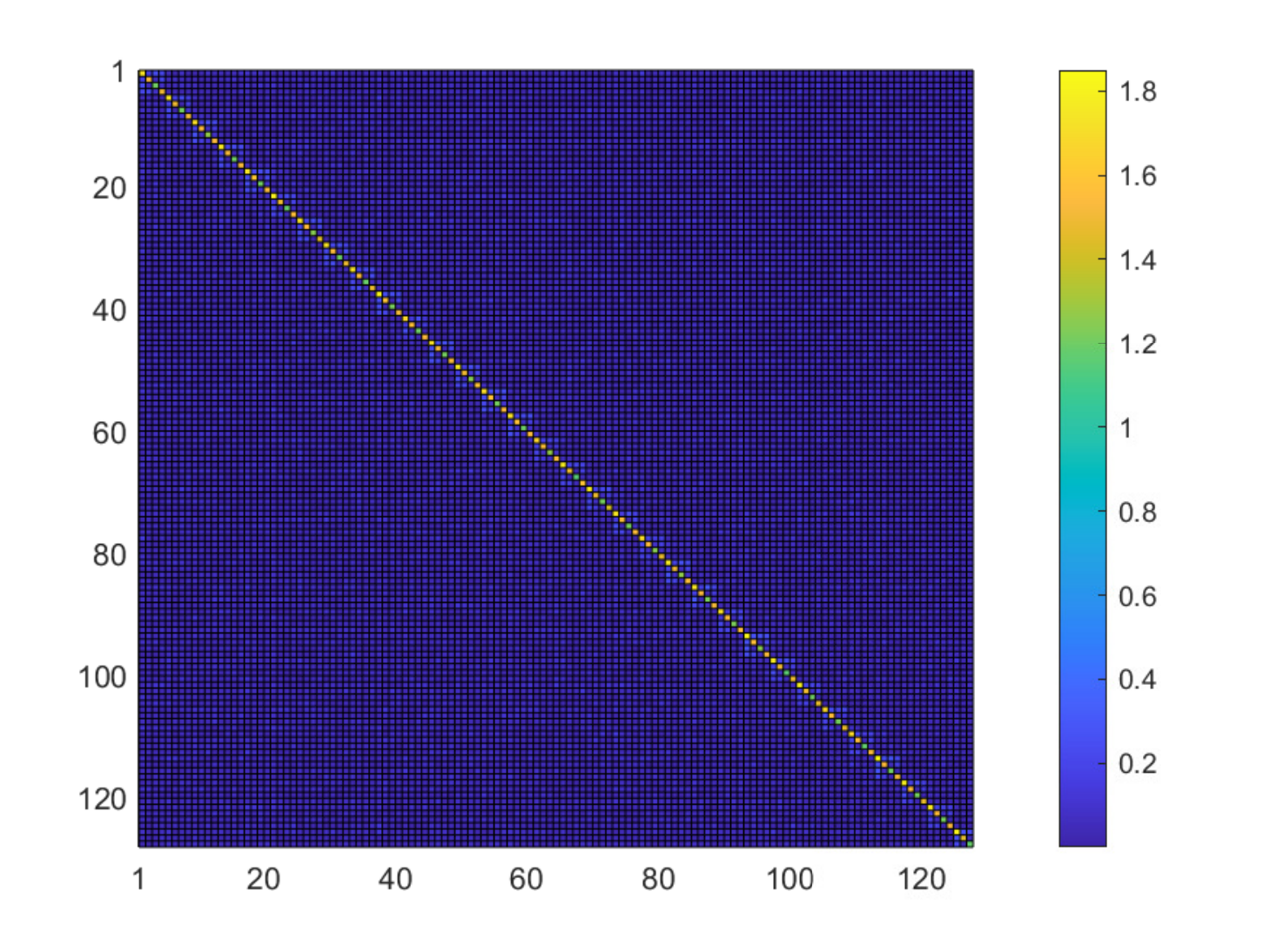}\\
        \caption{An example of $\mathbb{E}(\mathbf{s}_{\text{sup}}\mathbf{s}_{\text{sup}}^{\text{H}})$ with $M=16, N = 8$ and the codebook given in \cite{Huang2021}.}
        \label{Rx_SCMA}
    \end{minipage}
    \vspace{-7mm}
\end{figure}

Without loss of generality, we consider that the SCMA codewords are allocated on the DD domain plane along the delay direction in blocks of size $K \times 1$. It is worth pointing out that the proposed OTFS-SCMA detector can also work with other allocation schemes.

Thus, an OTFS frame comprises $\left\lfloor\frac{MN}{K}\right\rfloor$ SCMA codewords. And the transmitted DD domain symbol vector of the $j$-th user can be denoted by (with the assumption that $MN$ is exactly a multiple of $K$ in the following)
\begin{equation}
\mathbf{x}_j=[\mathbf{x}_{\text{SCMA},1}^{\text{T}},\mathbf{x}_{\text{SCMA},2}^{\text{T}},...,\mathbf{x}_{\text{SCMA},\frac{MN}{K}}^{\text{T}}]^{\text{T}},
\end{equation}
where $\mathbf{x}_{\text{SCMA},i}$ denotes the $i$-th SCMA codeword of user $j$.
Then, the transmitter superimposes the symbol vectors for all users in the DD domain and constructs the corresponding superimposed codeword vector $\mathbf{x}_{\text{sup}}$, i.e.
\begin{equation}
\mathbf{x}_{\text{sup}}=\sum_{j=1}^J \mathbf{x}_j.
\end{equation}
For the downlink user $j$, the received DD domain symbol vector can be modeled as
\begin{equation}
\mathbf{y}_j=\mathbf{H}_{j,\text{DD}}\mathbf{x}_{\text{sup}} + \tilde{\mathbf{n}}_j,
\end{equation}
where the subscript $j$ is used to distinguish the channel and the received signal for different users. Similarly, we can obtain the transmitted and received vectors of user $j$ in the time domain, i.e.
\begin{align}
\mathbf{s}_{\text{sup}} = (\mathbf{F}_N^{\text{H}} \otimes \mathbf{I}_M)\mathbf{x}_{\text{sup}}, \ \mathbf{r}_{j} = \mathbf{H}_{j,\text{T}}\mathbf{s}_{\text{sup}} + \mathbf{n}_j.
\end{align}

Fig. \ref{SCMA encoding and allocation} illustrates the SCMA encoding and allocation in the DD domain. For ease of exposition, we only illustrate the scenario for $M=8, N=4$. Nevertheless, the proposed algorithm can be extended to a larger OTFS-SCMA system since SCMA codewords are allocated in blocks.

\section{Joint OTFS-SCMA decoding with cross-domain detector}

\begin{figure}[t]
    \begin{minipage}[t]{0.48\linewidth}
        \centering
        \includegraphics[width=3in]{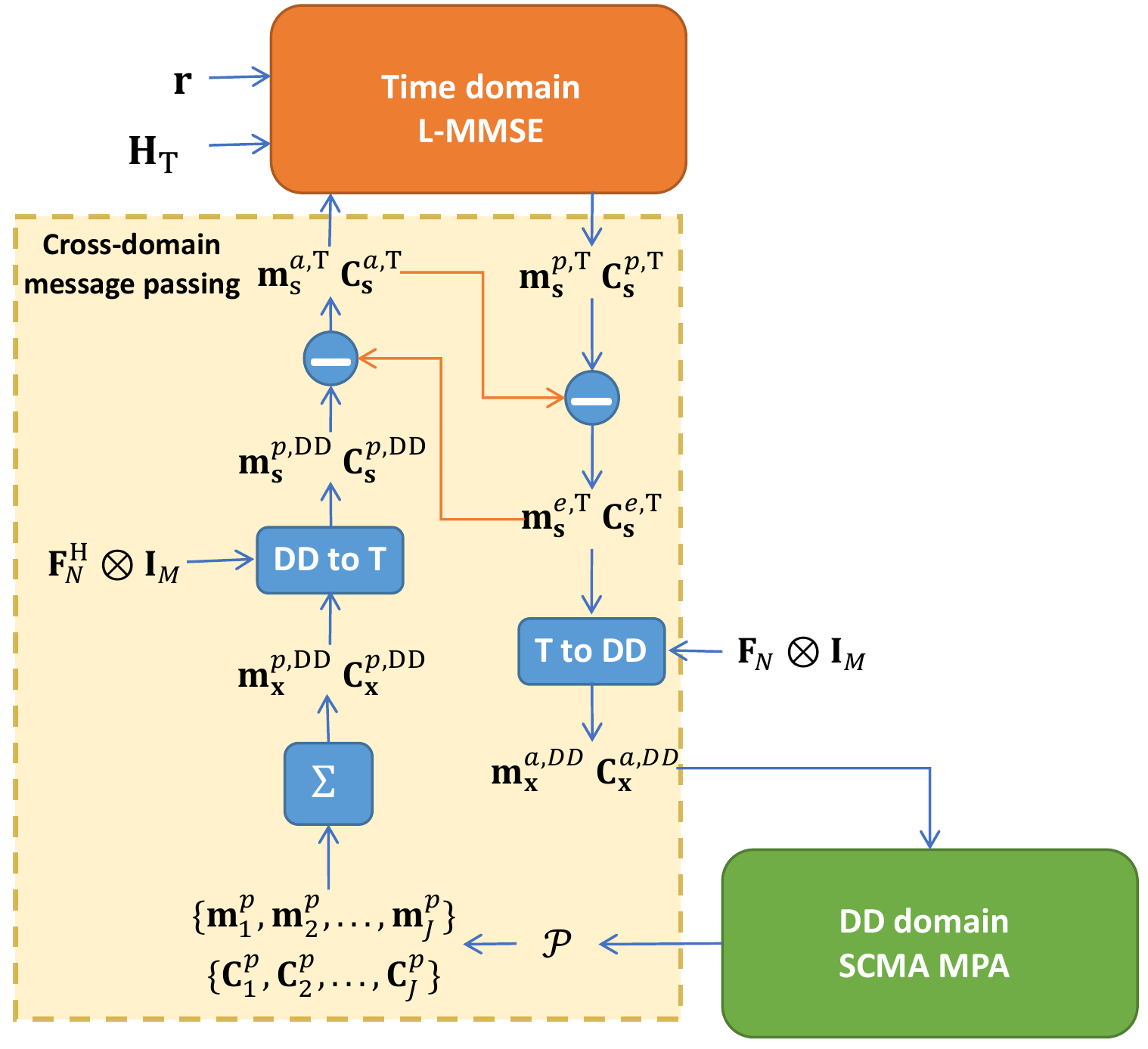}\\
        \caption{The single-layer structure of the proposed OTFS-SCMA cross-domain detector. The block ``T to DD" denotes the unitary transformation from the time domain to the DD domain, while ``DD to T" denotes the reverse. ``$\mathcal{P}$'' denotes the set of the \textit{a posterior} probabilities defined in (\ref{P_set}).
        ``$\Sigma$" denotes the operation of calculating the \textit{a posterior} information with respect to superimposed codewords in the DD domain, which is defined in (\ref{m_x_p}) and (\ref{C_x_p}).}
        \label{Detector}
    \end{minipage} \ 
    \begin{minipage}[t]{0.48\linewidth}
        \centering
        \includegraphics[width=3in]{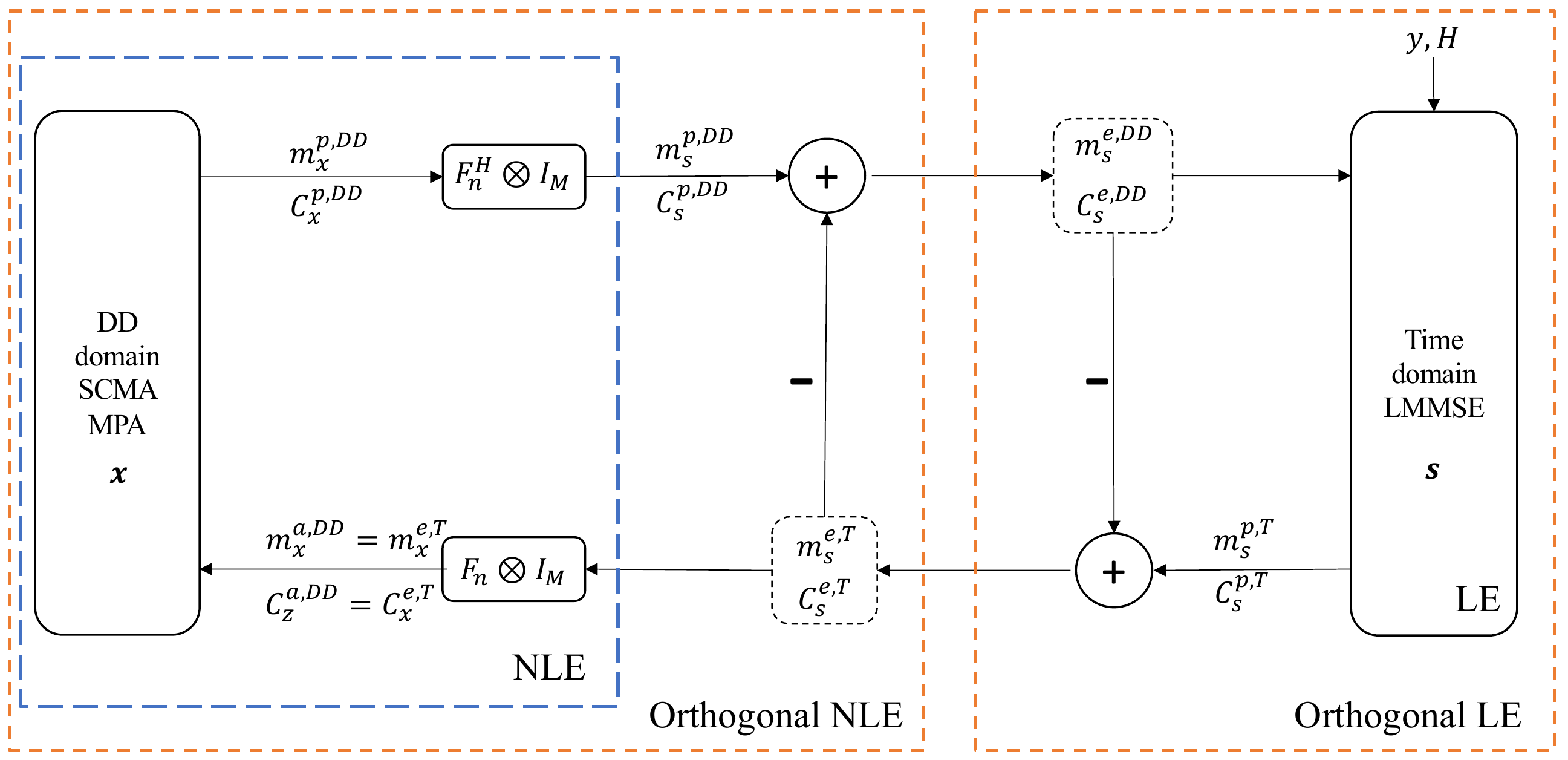}
        \caption{An illustration of cross-domain OTFS-SCMA detector in the view of orthogonal LE and NLE.}
        \label{OAMP_LE_NLE}
    \end{minipage} \quad
    \vspace{-0.6cm}
\end{figure}

{In this section, we propose a cross-domain OTFS-SCMA detector with a single-layer structure as shown in Fig. \ref{Detector}, which is performed locally in a downlink user.}
{In particular, we assume that the normalized superimposed codewords on each resource node are independently identically distributed (i.i.d.), i.e. $\mathbb{E}(\mathbf{x}_{\text{sup}}\mathbf{x}_{\text{sup}}^{\text{H}})=\mathbf{I}_{MN}$. Fig. \ref{Rx_SCMA} shows an example of $\mathbb{E}(\mathbf{x}_{\text{sup}}\mathbf{x}_{\text{sup}}^{\text{H}})$ by Monte Carlo simulation with $1000$ frames, indicating that this assumption is valid to a large extent.}
Based on the unitary transformation, the i.i.d. assumption of $\mathbf{s}_{\text{sup}}$ is also suitable, i.e., 
\begin{equation}
\mathbb{E}(\mathbf{s}_{\text{sup}}\mathbf{s}_{\text{sup}}^{\text{H}})=(\mathbf{F}^{\text{H}}_N\otimes \mathbf{I}_M)\mathbb{E}(\mathbf{x_{\text{sup}}}\mathbf{x_{\text{sup}}}^{\text{H}})(\mathbf{F}_N\otimes \mathbf{I}_M)=\mathbf{I}_{MN}.
\vspace{-2mm} 
\end{equation}
Also, we assume that the entries in $\mathbf{s}_{\text{sup}}$ is Gaussian variables due to the spreading effect of ISFFT.

Inspired by the priciple of errors orthogonality of OAMP \cite{Ma2017, Liu2022}, we design a cross-domain OTFS-SCMA detector consisting of a linear estimator (LE) in the time domain and a non-linear estimator (NLE) in the DD domain with unitary transformation. 
To satisfy the orthogonality of errors, we develop a cross-domain message passing algorithm as shown in the yellow area of Fig. \ref{Detector}. In the view of orthogonal LE and NLE, the proposed cross-domain detector can also be illustrated in Fig. \ref{OAMP_LE_NLE}. Denote the NLE and orthogonal NLE in Fig. \ref{OAMP_LE_NLE} by $\hat{\phi}(\mathbf{x})$ and $\phi(\mathbf{x})$, respectively, and the LE and orthogonal LE by $\hat{\gamma}(\mathbf{x})$ and $\gamma(\mathbf{x})$, respectively. Thus, the cross-domain OTFS-SCMA detector can be written as\footnote{The details of the derivation of the (\ref{orth_gamma}) and (\ref{orth_phi}) will be described in the following subsections. \vspace{-2mm}}
\begin{align}
    \gamma(\mathbf{x}^{\phi\rightarrow\gamma}) &= \left( \mathbf{C}^{e,T}_{s} \left( \mathbf{C}_{s}^{p,T} \right)^{-1} \right) \hat{\gamma}(\mathbf{x}^{\phi\rightarrow\gamma}) - \left( \mathbf{C}^{e,T}_{s} \left( \mathbf{C}_{s}^{a,T} \right)^{-1} \right) \mathbf{x}^{\phi\rightarrow\gamma}, \label{orth_gamma} \\
    \phi(\mathbf{x}^{\gamma\rightarrow\phi}) &= \left( \mathbf{C}^{e,DD}_{s} \left( \mathbf{C}_{s}^{p,DD} \right)^{-1} \right) \hat{\phi}(\mathbf{x}^{\gamma\rightarrow\phi}) - \left( \mathbf{C}^{e,DD}_{s} \left( \mathbf{C}_{s}^{a,DD} \right)^{-1} \right) \mathbf{x}^{\gamma\rightarrow\phi}. \label{orth_phi}
\end{align}
We can observe that the above iterative process is indeed a type of expectation propagation (EP). Furthermore, by \cite{Liu2022}, the EP and OAMP are equivalent when locally optimal prototypes are employed, e.g., LMMSE estimator and SCMA MPA decoder. Therefore, the errors between the time domain LMMSE and the DD domain SCMA MPA decoder are orthogonal with the help of cross-domain message passing, thus giving rise to enhanced convergence in iterative decoding.
{Note that the proposed OTFS-SCMA iterative detector based on error orthogonality is fundamentally different from the turbo-based iterative detectors, e.g., turbo-based LDPC SCMA detector \cite{Xiao2015LDPCSCMA} and the joint polar-SCMA detector \cite{Li2022TVT}, which require independent input-output errors of each local estimator. Moreover, the proposed method exploits the unitary transform that replaces the conventional interleaver in turbo-based detectors.}



\subsection{Time domain L-MMSE equalization}

The conventional L-MMSE equalizer is employed to estimate the time domain OTFS-SCMA superimposed symbol vector $\mathbf{s}_{\text{sup}}$. The received time domain symbol vector $\mathbf{r}$ and the time domain channel matrix $\mathbf{H}_{\text{T}}$ with the aid of the \textit{a priori} mean $\mathbf{m}_\mathbf{s}^{a,\text{T}}$ and the \textit{a priori} covariance matrix $\mathbf{C}_\mathbf{s}^{a,\text{T}}$, are fed to the L-MMSE equalizer to calculate the estimation matrix and return a rough estimate. Note that due to the i.i.d. assumption, $\mathbf{C}_\mathbf{s}^{a,\text{T}}$ is diagonal matrix and thus initialized as an identity matrix $\mathbf{I}_{MN}$. And the \textit{a priori} mean $\mathbf{m}_\mathbf{s}^{a,\text{T}}$ is initialized as zeros. 
 Based on the L-MMSE estimation matrix, the \textit{a posteriori} estimation mean $\mathbf{m}_\mathbf{s}^{p,\text{T}}$ and covariance matrix $\mathbf{C}_\mathbf{s}^{p,\text{T}}$ of $\mathbf{s}_{\text{sup}}$ are given by
\begin{align}
\mathbf{m}_\mathbf{s}^{p,\text{T}} &= \mathbf{m}_\mathbf{s}^{a,\text{T}} + \mathbf{W}_{\text{MMSE}}(\mathbf{r}-\mathbf{H}_{\text{T}}\mathbf{m}_\mathbf{s}^{a,\text{T}}), \label{LMMSE_m} \\
\mathbf{C}_\mathbf{s}^{p,\text{T}} &= \mathbf{C}_\mathbf{s}^{a,\text{T}} - \mathbf{W}_{\text{MMSE}}\mathbf{H}_{\text{T}}\mathbf{C}_\mathbf{s}^{a,\text{T}}, \label{LMMSE_C} 
\end{align}
where $\mathbf{W}_{\text{MMSE}}$ is the L-MMSE estimation matrix, and it is defined as \cite{Kay1993}
\begin{equation}
\mathbf{W}_{\text{MMSE}}=\mathbf{C}_\mathbf{s}^{a,\text{T}}\mathbf{H}_{\text{T}}^{\text{H}}\left(\mathbf{H}_{\text{T}} \mathbf{C}_\mathbf{s}^{a,\text{T}} \mathbf{H}_{\text{T}}^{\text{H}} + N_0\mathbf{I}_{MN} \right)^{-1}.
\label{LMMSE_W} 
\end{equation}
Note that the non-diagonal entries in $\mathbf{C}_\mathbf{s}^{p,\text{T}}$ are treated as zeros since only diagonal entries are of interest according to the i.i.d. assumption.

Having the \textit{a posteriori} mean $\mathbf{m}_\mathbf{s}^{p,\text{T}}$ and covariance matrix $\mathbf{C}_\mathbf{s}^{p,\text{T}}$ in hand, the extrinsic information for time domain can be calculated and passed to DD domain SCMA decoder via the unitary transformation, which will be discussed in Subsection \uppercase\expandafter{\romannumeral3}-C.

\subsection{MPA based SCMA decoder}

The \textit{a priori} mean $\mathbf{m}_\mathbf{x}^{a,\text{DD}}$ and covariance matrix $\mathbf{C}_\mathbf{x}^{a,\text{DD}}$ calculated by the extrinsic information from the time domain equalizer are fed to the SCMA decoder.
Since the \textit{a priori} mean $\mathbf{m}_\mathbf{x}^{a,\text{DD}}$ can be viewed as rough estimates of the superimposed codewords, the SCMA detection problem can be formulated in the DD domain by
\begin{equation}
\mathbf{m}_{\mathbf{x}}^{a,\text{DD}}=\mathbf{x}_{\text{sup}}+\hat{\mathbf{n}},
\label{detection_problem_DD}
\end{equation}
{where $\hat{\mathbf{n}}$ is modeled as a white Gaussian noise sample vector characterizing the uncertainty of the time domain estimates with zero mean and a covariance matrix $\mathbf{C}_{\mathbf{x}}^{a,\text{DD}}$ \cite{Yuan2014,Liu2022}.}

Note that $\mathbf{x}_{\text{sup}}$ is the summation of $\mathbf{x}_j,\forall j$ and $\mathbf{x}_j$ is stacked by $MN/K$ SCMA codewords. Therefore, we can conduct the SCMA decoding in a codeword-by-codeword manner. In particular, we define that $X_{i,j}\in\mathbb{C}^{K}, 0\le i \le \frac{MN}{K}-1$ is the $i$-th codeword in $\mathbf{x}_j$, $X_{i,\text{sup}}\in\mathbb{C}^{K}$ is the $i$-th superimposed codeword in $\mathbf{x}_{\text{sup}}$, i.e., $X_{i,\text{sup}}=\sum_j X_{i,j}$,  and $\mathbf{m}_i\in\mathbb{C}^{K}$ is the $i$-th roughly estimated superimposed codeword in $\mathbf{m}_{\mathbf{x}}^{a,\text{DD}}$. In a codeword-by-codeword manner, Eq. (\ref{detection_problem_DD}) can be rewritten as
\begin{equation}
\mathbf{m}_i=X_{i,\text{sup}}+\hat{\mathbf{n}}_i, \quad \forall i,
\end{equation}
where $\hat{\mathbf{n}}_i$ is the Gaussian uncertainty term corresponding to the $i$-th codeword.
For an SCMA decoder, the maximum \textit{a posteriori} (MAP) detection of the $i$-th codeword is given by
\begin{equation}
\{\hat{X}_{i,j}\}_{1\le j\le J}=\arg\max_{X_{i,j}\in\mathcal{X}_j,\forall j}{p(X_{i,\text{sup}}|\mathbf{m}_i)}.
\end{equation}
However, the complexity of MAP detection is $\mathcal{O}(M_{\text{mod}}^J)$, which is extremely high when the number of users is large. With the aid of the factor graph (e.g., Fig. \ref{Factor graph}), the MPA decoder can be employed to decode SCMA codewords and approach the error performance of the MAP detector \cite{Nikopour2013, Hoshyar2008}. Specifically, the decoding can be divided into three steps.

\subsubsection{Initialization}

Given a factor graph $\mathbf{F}^{\text{ind}}$ (see (\ref{factor graph matrix})), the position sets of $\mathbf{F}^{\text{ind}}$ are defined as $\zeta_j=\{k|\mathbf{F}^{\text{ind}}[k,j]=1,\forall k\}$ for $\forall j$, and $\xi_k=\{j|\mathbf{F}^{\text{ind}}[k,j]=1,\forall j\}$ for $\forall k$. Then, for each FN, i.e., for the $k$-th resource node, the likelihood function is initialized as
\begin{equation}
    \begin{aligned}
   f\left(\mathbf{m}_i[k]\big|X_{i,\text{sup}}\right) = \exp\Bigg\{-\frac{1}{2 \sigma_{\mathbf{\hat{\mathbf{n}}}}^2} \Big|\mathbf{m}_i[k] -  \sum_{j\in\xi_k} X_{i,j}[k]\Big|^2 \Bigg\}, \  \forall \ X_{i,j}\in\mathcal{X}_{j},
    \end{aligned}
    \label{init_f}
\end{equation}
where $\sigma_{\mathbf{\hat{\mathbf{n}}}}^2 = \frac{1}{MN}\text{Tr}(\mathbf{C}_{\mathbf{x}}^{a,\text{DD}})$ due to the i.i.d. assumption in the DD domain.

The initial message passed from VNs to the $k$-th FN is set to be $\frac{1}{M_{\text{mod}}}$, i.e. 
\begin{equation}
    p_a=\eta_{j\rightarrow k}^0(X_{i,j})=\frac{1}{M_{\text{mod}}}, 
    \label{init_p}
\end{equation}
due to the assumption of equal prior probability for each codeword.

\subsubsection{Exchanges extrinsic information between VNs and FNs iteratively}

In the $q$-th iteration, the message passing is given by
\begin{equation}
    \begin{aligned}
        \eta^q_{k\rightarrow j}(X_{i,j}) 
        =\sum_{\tilde{X}_{i,\text{sup}}:\tilde{X}_{i,j}=X_{i,j}}\left(f(\mathbf{m}_i[k]|\tilde{X}_{i,\text{sup}})\prod_{\tilde{j}\in\xi_{k}\backslash j}\eta_{\tilde{j}\rightarrow k}^{q-1}(\tilde{X}_{i,\tilde{j}}) \right),
    \end{aligned}
    \label{FN2VN}
\end{equation}
and
\begin{equation}
    \eta^q_{j\rightarrow k}(X_{i,j})  = \text{normalize}\left(p_a\prod_{\tilde{k}\in \zeta_{j}\backslash k} \eta^{q}_{\tilde{k}\rightarrow j}(X_{i,j})\right),
    \label{VN2FN}
\end{equation}
where $\tilde{X}_{i,\text{sup}}:\tilde{X}_{i,j}=X_{i,j}$ denotes all possible superimposed codewords with $\tilde{X}_{i,j}=X_{i,j}$, $\tilde{j}\in\xi_k\backslash j$ denotes that all VNs carried by FN $k$ expect for the $j$-th VN, and $\tilde{k}\in\zeta_j\backslash k$ denotes that all FNs connected to $j$-th VN expect for the $k$-th FN.

\subsubsection{Selection of codewords}

After $I_q$ iterative computing, the $i$-th codeword for each $j$ is estimated as
\begin{equation}
   \hat{X}_{i,j}=\arg\max_{X_{i,j}\in \mathcal{X}_j}\left\{\prod_{k\in\zeta_j}\eta^{I_q}_{k\rightarrow j}(X_{i,j})\right\}.
   \label{sel_cw}
\end{equation}

Note that the term $P\left(X_{i,j}=(\mathcal{X}_j)_m |\mathbf{m}_i\right)=p_a\prod_{k\in\zeta_j}\eta^{I_q}_{k\rightarrow j}(X_{i,j})$ is the \textit{a posteriori} probability of arbitrary codeword of the $j$-th user $X_{i,j} \in \mathcal{X}_j$, where $(\mathcal{X}_j)_m$ is the $m$-th codeword of the codebook $\mathcal{X}_j$ with size $M_{\text{mod}}$.
Therefore, we can formulate the \textit{a posteriori} probability set consisting of all $P(X_{i,j})$ for $\forall i,j$, i.e.
\begin{equation} 
    \begin{aligned}
    \mathcal{P} = \Bigg\{ & P\left(X_{i,j}=(\mathcal{X}_j)_m \big| \mathbf{m}_i\right), 
     1\le j\le J,\  0\le i\le MN/K,\  1\le m\le M_{\text{mod}}\Bigg\}.
    \end{aligned}
    \label{P_set}
\end{equation}
The cardinality of $\mathcal{P}$ is $|\mathcal{P}|=\frac{MNJM_{\text{mod}}}{K}$. 

Then the  \textit{a posteriori} mean $\mathbf{m}_{j,X_{i,j}}^p$ and covariance matrix $\mathbf{C}_{j,X_{i,j}}^p$ of the $j$-th user's $i$-th codeword can be calculated as
\begin{equation}
\mathbf{m}_{j,X_{i,j}}^p =\sum_{\beta\in\mathcal{X}_j}\beta P\left(X_{i,j}=\beta|\mathbf{m}_i\right),
\label{m_j_p}
\end{equation}
and
\begin{equation}
    \begin{aligned}
\mathbf{C}_{j,X_{i,j}}^p = \sum_{\beta\in\mathcal{X}_j} \Bigg\{  \left(\beta-\mathbf{m}_{j,X_{i,j}}^p\right)\left(\beta-\mathbf{m}_{j,X_{i,j}}^p\right)^H 
     P\left(X_{i,j}=\beta|\mathbf{m}_i\right)\Bigg\}.
    \end{aligned}
\end{equation}
Due to the i.i.d. assumption, the $\mathbf{C}^p_{j,X_{i,j}}$ of $X_{i,j}$ is a diagonal matrix whose $k$-th element in the main diagonal is the \textit{a posteriori} variance of $X_{i,j}[k]$ which is given by
\begin{equation}
\begin{aligned}
\mathbf{C}^{p}_{j,X_{i,j}}[k,k] 
&= \mathbb{E} \left[ \left|X_{i,j}[k]-\mathbb{E}\left[X_{i,j}[k]|\mathbf{m}^p_{j,X_{i,j}}\right]\right|^2\right] \\
&= \sum_{\beta\in\mathcal{X}_j} |X_{i,j}[k]|^2 P(X_{i,j}[k]=\beta[k]|\mathbf{m}_i)-|\mathbf{m}^p_{j,X_{i,j}}[k]|^2.
\end{aligned}
\label{C_j_p}
\end{equation}
It is worth noting that, there are only $d_v$ non-zero elements in the main diagonal of $\mathbf{C}^p_{j,X_{i,j}}$ due to the sparsity of the SCMA codebooks, and $\mathbf{C}^p_{j,X_{i,j}}$ is rank-deficient.

Then the  \textit{a posteriori} mean $\mathbf{m}_{j}^p$ and covariance matrix $\mathbf{C}_{j}^p$ of user $j$ in the DD domain are formulated by stacking all the $\mathbf{m}_{j,X_{i,j}}^p$ and $\mathbf{C}_{j,X_{i,j}}^p$ for all $i$, in the order of the SCMA allocation scheme in the DD domain grid.

\subsection{Cross-domain message passing}

In this subsection, we derive the extrinsic information passing between the time domain and the DD domain, i.e., the yellow part in Fig. \ref{Detector}. 

\subsubsection{From time domain to DD domain}

First, we calculate the extrinsic mean and covariance matrix from the time domain equalizer by
\begin{align}
&\mathbf{C}^{e,\text{T}}_\mathbf{s}=\left(\left(\mathbf{C}_\mathbf{s}^{p,\text{T}}\right)^{-1}-\left(\mathbf{C}_\mathbf{s}^{a,\text{T}}\right)^{-1} \right)^{-1}, \label{ext_T_C}\\
&\mathbf{m}_\mathbf{s}^{e,\text{T}}=\mathbf{C}_\mathbf{s}^{e,\text{T}}\left( \left( \mathbf{C}_\mathbf{s}^{p,\text{T}} \right)^{-1} \mathbf{m}_\mathbf{s}^{p,\text{T}} - \left( \mathbf{C}_\mathbf{s}^{a,\text{T}} \right)^{-1}\mathbf{m}_\mathbf{s}^{a,\text{T}} \right). \label{ext_T_m}
\end{align}
{Note that (\ref{ext_T_m}) is equivalent to (\ref{orth_gamma}).}
With the aid of the unitary transformation from the time domain to the DD domain ``$\mathbf{F}_N\otimes \mathbf{I}_M$'', the \textit{a priori} mean and covariance matrix of $\mathbf{x}$ are given by
\begin{align}
&\mathbf{m}_{\mathbf{x}}^{a,\text{DD}}=\mathbf{m}_{\mathbf{x}}^{e,\text{T}}=(\mathbf{F}_N\otimes \mathbf{I}_M)\mathbf{m}_\mathbf{s}^{e,\text{T}}, \label{ext_T2DD_m}\\
&\mathbf{C}_{\mathbf{x}}^{a,\text{DD}}=\mathbf{C}_{\mathbf{x}}^{e,\text{T}}=(\mathbf{F}_N\otimes \mathbf{I}_M)\mathbf{C}_\mathbf{s}^{e,\text{T}}(\mathbf{F}_N^\text{H}\otimes \mathbf{I}_M).
\label{ext_T2DD_C}
\end{align}
Note that for the large $MN$ setting, the diagonal entries of the diagonal matrix $\mathbf{C}_\mathbf{s}^{e,\text{T}}$ tends to be the same value due to the strong law of large numbers, and thus the matrix $\mathbf{C}_{\mathbf{x}}^{a,\text{DD}}$ tends to be diagonal.

\subsubsection{From DD domain to time domain}
After MPA decoding, we have the \textit{a posteriori} mean $\mathbf{m}_j^p$ and covariance matrix $\mathbf{C}_j^p$ in hand. However, the time domain OTFS symbol estimation is carried out with respect to the superimposed codewords i.e. $\mathbf{s}_{\text{sup}}$. We have to reconstruct the superimposed codewords by summing the decoded SCMA codewords for each user. Hence, the corresponding \textit{a posteriori} mean and covariance matrix of $\mathbf{x}$ in DD domain are given by
\begin{equation}
\mathbf{m}_{\mathbf{x}}^{p,\text{DD}}=\sum_{j=1}^J{\mathbf{m}_j^p}, \label{m_x_p}
\vspace{-0.4cm}
\end{equation}
and
\begin{equation}
    \mathbf{C}_{\mathbf{x}}^{p,\text{DD}}=\left(\sum_{j=1}^J\left(\mathbf{C}_j^p\right)^{-1}\right)^{-1}.
\end{equation}
Nevertheless, the $j$-th covariance matrix $\mathbf{C}_j^p$ is rank-deficient, and thus may be non-invertable. Thanks to the i.i.d. assumption, $\mathbf{C}_{\mathbf{x}}^{p,\text{DD}}$ is a diagonal matrix and can be obtained by 
\begin{equation}
\mathbf{C}_{\mathbf{x}}^{p,\text{DD}}[i,i] = \left(\sum_{j\in \xi_k} {\frac{1}{\mathbf{C}_j^p[i,i]}} \right)^{-1},
\label{C_x_p}
\end{equation}
where $k=(i\bmod{M})+1$ and $0\le i\ \le MN-1$.

Then, we convert $\mathbf{m}_{\mathbf{x}}^{p,\text{DD}}$ and $\mathbf{C}_{\mathbf{x}}^{p,\text{DD}}$ to the \textit{a posteriori} mean and covariance matrix of the time domain OTFS signal $\mathbf{s}$ by
\begin{align}
\mathbf{m}_\mathbf{s}^{p,\text{DD}} &= \left(\mathbf{F}_N^\text{H} \otimes \mathbf{I}_M \right)\mathbf{m}_{\mathbf{x}}^{p,\text{DD}}, \label{p_DD2T_m} \\
\mathbf{C}_\mathbf{s}^{p,\text{DD}} &= (\mathbf{F}_N^\text{H}\otimes \mathbf{I}_M)\mathbf{C}_{\mathbf{x}}^{p,\text{DD}}(\mathbf{F}_N\otimes \mathbf{I}_M). \label{p_DD2T_C}
\end{align}
Similar to (35) and (36), the extrinsic information of $\mathbf{s}$ in terms of the mean and covariance matrix is given by
\begin{equation}
\mathbf{C}^{a,\text{T}}_\mathbf{s}=\mathbf{C}^{e,\text{DD}}_\mathbf{s}=\left(\left(\mathbf{C}_\mathbf{s}^{p,\text{DD}}\right)^{-1}-\left(\mathbf{C}_\mathbf{s}^{e,\text{T}}\right)^{-1} \right)^{-1},
\label{C_a_T}
\end{equation}
and
\begin{equation}
    \begin{aligned}
\mathbf{m}^{a,\text{T}}_\mathbf{s} = \mathbf{m}_\mathbf{s}^{e,\text{DD}} = \mathbf{C}_\mathbf{s}^{e,\text{DD}}\left( \left( \mathbf{C}_\mathbf{s}^{p,\text{DD}} \right)^{-1} \mathbf{m}_\mathbf{s}^{p,\text{DD}} - \left( \mathbf{C}_\mathbf{s}^{e,\text{T}} \right)^{-1}\mathbf{m}_\mathbf{s}^{e,\text{T}} \right).
    \end{aligned}
    \label{m_a_T}
\end{equation}
{Similarly, (\ref{m_a_T}) is equivalent to the (\ref{orth_phi}).}

After a number of cross-domain message passing iterations, e.g. $L_{\text{max}}$, or $\frac{1}{MN}\text{Tr}(\mathbf{C}_{\mathbf{s}}^{a,\text{T}})<10^{-3}$, the detector returns the detected SCMA codewords by (\ref{sel_cw}).

\subsection{Complexity Analysis}
The complexity of the time domain L-MMSE equalizer is dominant by the matrix inverse in (\ref{LMMSE_W}), whose complexity order is $\mathcal{O}\left((MN)^3\right)$. 
In the SCMA decoding part, the complexity of the conventional MPA decoder is given as $\mathcal{O}\left(I_qMN(M_{\text{mod}})^{d_c}d_c^2\right)$ \cite{Liu2021}, where $I_q$ is the number of MPA iterations and $d_c$ is the number of non-zero entries in each row of indicator matrix $\mathbf{F}^{\text{ind}}$ in (\ref{factor graph matrix}). 
During cross-domain message passing, the unitary transformation with respect to $\mathbf{F}_N\otimes \mathbf{I}_M$ and $\mathbf{F}_N^{\text{H}}\otimes \mathbf{I}_M$ can be efficiently calculated by fast Fourier transform (FFT) and inverse FFT with complexity $\mathcal{O}\left(MN\log N\right)$. And the computational complexity of covariance matrix inverse in (\ref{ext_T_C}), (\ref{ext_T_m}), (\ref{C_a_T}), and (\ref{m_a_T}) can be reduced by only calculating the diagonal entries, according to the i.i.d. assumption, having the complexity order of $\mathcal{O}(MN)$.
The total detection complexity of the proposed OTFS-SCMA detector per iteration can be given by $\mathcal{O}\left((MN)^3+I_qMN(M_{\text{mod}})^{d_c}d_c^2+MN\log N+MN\right)$. 
The complexity is high due to the high computational complexity of matrix inverse in the L-MMSE equalizer. However, this complexity can be further reduced by adopting some low-complexity estimation/equalization algorithms that approximate the L-MMSE performance. For instance, the MMSE equalizers with log-linear order of complexity proposed in \cite{Tiwari2019} and \cite{Surabhi2020} can simply replace the time domain L-MMSE equalizer in our proposed OTFS-SCMA detector.

\subsection{Discussion}
The proposed cross-domain detection leads to some advantages of OTFS-SCMA compared to the conventional OFDM-SCMA. 
{The proposed cross-domain detection leads to some advantages of OTFS-SCMA compared to the conventional OFDM-SCMA. 
First, the OTFS-SCMA system inherits the advantages of OTFS, e.g., potential ability to exploit full diversity gain, resilience to narrowband interference, low peak-to-average power ratio (PAPR), etc., which are not available in OFDM-SCMA systems \cite{Hadani2017,Wei2021}.} 
On the other hand, the error performance of SCMA systems mainly depends on the minimum Euclidean distance (MED) and the minimum product distance (MPD) between superimposed codewords. In conventional OFDM-SCMA systems, a larger MED leads to better BER performance over Gaussian channels, while a larger MPD contributes to improve performance over Rayleigh fading channels \cite{Boutros1998,Vameghestahbanati2019}. However, the design of SCMA codebook with both large MED and MPD is difficult \cite{Liu2021,Taherzadeh2014}.
The cross-domain detection for OTFS-SCMA systems is promising to solve this problem. Based on the observation of (\ref{detection_problem_DD}), the inputs of SCMA decoder in the DD domain may be roughly regarded as the signal over AWGN channels. Thus, the error performance of the SCMA decoder in the DD domain mainly depends on the MED of superimposed codewords instead of MPD. This observation gives us an insight that well-designed SCMA codebooks for AWGN channel with large MED work well in OTFS-SCMA transmissions. We validate this observation by our numerical results in Section IV.

\begin{figure}[t]
    \begin{minipage}[t]{0.48\linewidth}
      \centering
      \includegraphics[width=3in]{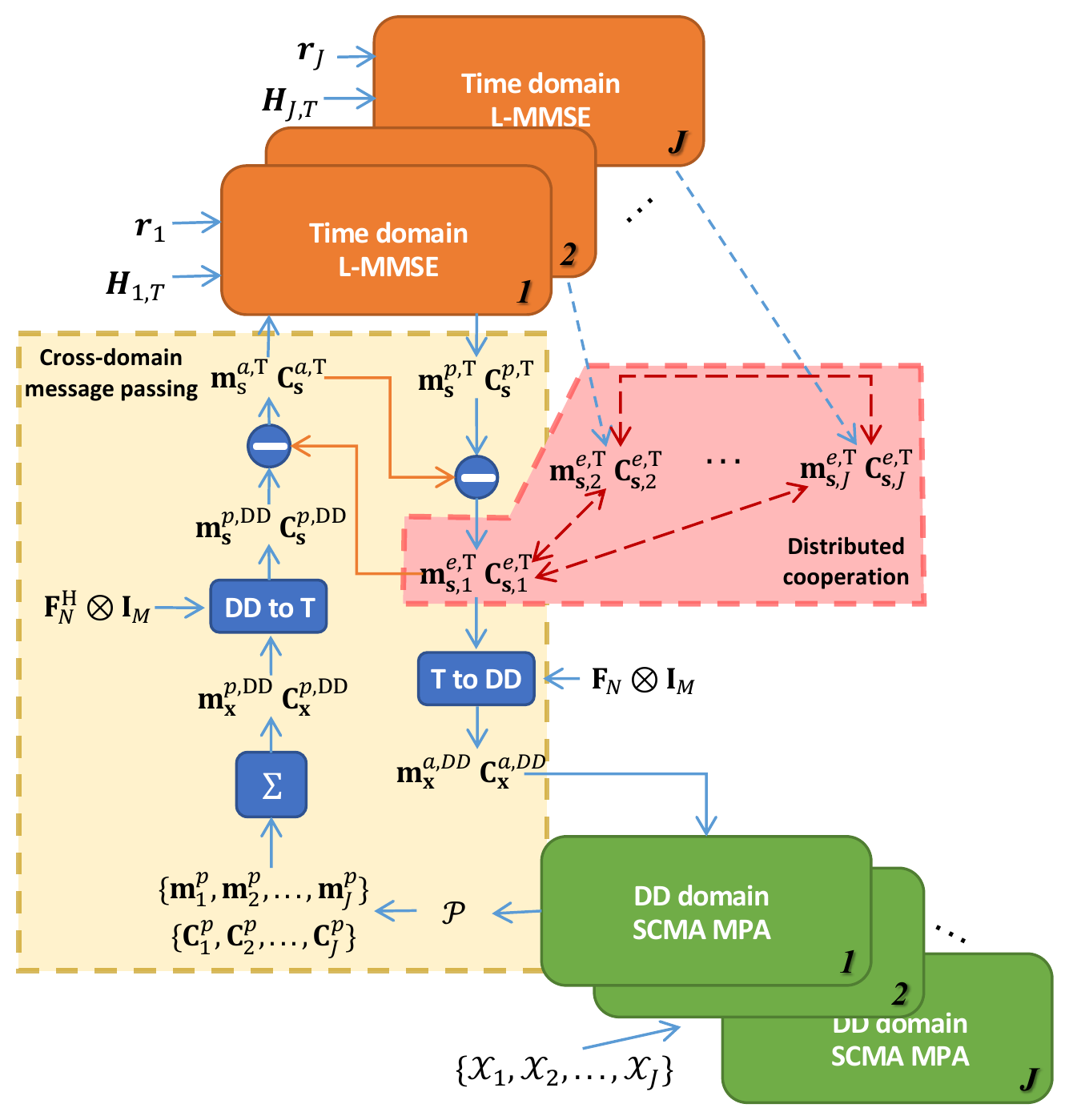}\\
      \caption{The multi-layer structure of the proposed OTFS-SCMA cross-domain detector. }
      \label{Multi-layer Detector}
    \end{minipage} \quad \  
    \begin{minipage}[b]{0.48\linewidth}
      \centering
      \includegraphics[width=2.8in]{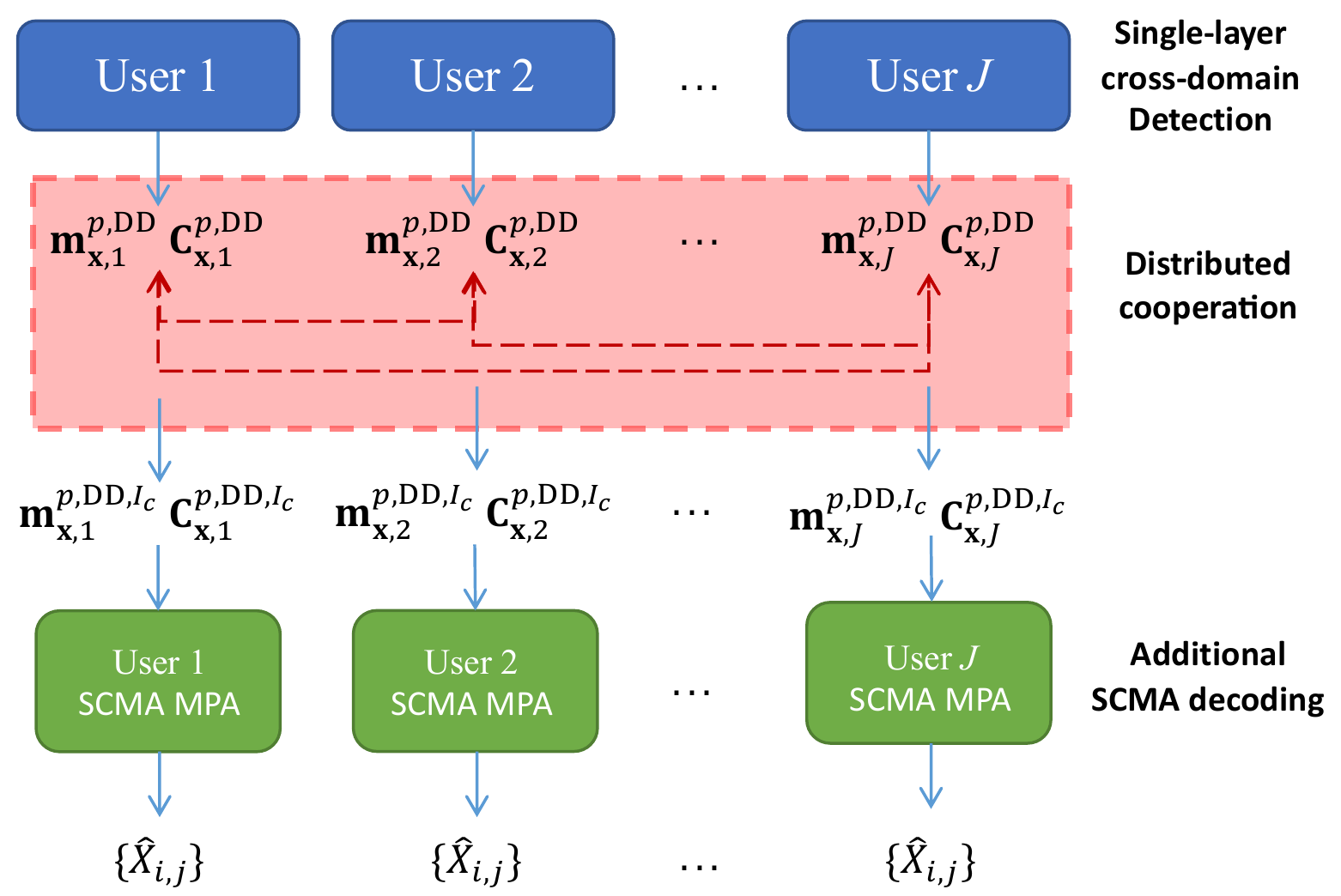}\\
      \caption{An illustration of the proposed separate cross-domain and DCD. The block ``User $j$'' denotes the local single-layer cross-domain detection of the user $j$, which is illustrated in Fig. \ref{Detector}. $\{\hat{X}_{i,j}\}$ is the set of decoded SCMA codewords that is defined by (\ref{sel_cw}).}
      \label{Separated structure}
    \end{minipage}
    \vspace{-0.5cm}
\end{figure}


\section{Joint cross-domain and distributed cooperative detection}

In this section, we consider the multi-layer structure of the proposed OTFS-SCMA detector shown in Fig. \ref{Multi-layer Detector}. By introducing a cooperative network, the single-layer structure described in Section \uppercase\expandafter{\romannumeral3} can be extended into a multi-layer structure to achieve large user diversity gains from other downlink users compared to the previous single-layer detection. In this paper, we assume that there exists a simple and efficient communication scheme that supports proximity users to share information with each other, e.g. Cellular V2X \cite{Wang2018}, which does not consume many wireless resources.

Since the received signal for downlink users experience different channels, we consider that all downlink users share their extrinsic information of time domain to nearby users to exploit the diversity gains.
Specifically, in each cross-domain message passing iteration, the extrinsic information from time domain equalizer of the $j$-th user is broadcasted to several nearby users depending on a given distance threshold or a given neighboring set. At the same time, the $j$-th user updates its local extrinsic information based on the received extrinsic information from nearby users.

Let $\mathcal{S}_j$ be the neighboring set of user $j$, i.e., users in $\mathcal{S}_j$ can receive the information from user $j$. {Then our goal is to obtain the product of all users' messages distributively, i.e., to agree on the global message at each user with only local processing and cooperation with nearby users.}

\subsection{Belief consensus-based method}
The belief consensus method is efficient to compute the product of several local functions over the same variable distributively \cite{Xiao2005}. With the assumption of Gaussian messages, users are able to exchange parameters of the messages instead of the distribution, i.e., we can only broadcast means and covariance matrices. In this case, the $j$-th user updates its local belief according to standard belief consensus recursion, i.e.,
\begin{equation}
    \mathbf{\theta}_{i,j}^{c+1} = \gamma_{jj}\mathbf{\theta}_{i,j}^c+\sum_{g\in \mathcal{S}_j}\gamma_{jg}\mathbf{\theta}_{i,g}^c,\quad 0\le i\le MN-1,
    \label{consensus_perfect}
\end{equation}
where the superscript $c$ denotes the index of consensus iterations and $\gamma$ is the update rate defined as \cite{Xiao2005}
\begin{equation}
    \gamma_{jg}=\gamma_{gj}=
\begin{cases}
1/\max(|\mathcal{S}_j|,|\mathcal{S}_g|),\quad & \text{for} \ j\ne g, \\
1-\sum_{j'\in \mathcal{S}_g}\gamma_{j'g}, & \text{for} \ j=g.
\end{cases}
\label{update_rate}
\end{equation}

For the proposed joint cross-domain and DCD algorithm, the belief consensus iteration is integrated in cross-domain message passing process. The local information to be shared is constructed by the extrinsic mean and covariance matrix from the time domain equalizer, i.e.
 \begin{equation}
    \mathbf{\theta}_{i,j}^{c}=\left[\mathbf{m}_{\mathbf{s},j}^{e,\text{T},c}[i]/\mathbf{C}_{\mathbf{s},j}^{e,\text{T},c}[i,i], 1/\mathbf{C}_{\mathbf{s},j}^{e,\text{T},c}[i,i]\right]^T,
 \end{equation}
where $\mathbf{m}_{\mathbf{s},j}^{e,\text{T},c}$ and $\mathbf{C}_{\mathbf{s},j}^{e,\text{T},c}$ denote the extrinsic mean and covariance matrix of user $j$ in the $c$-th belief consensus iteration, respectively.

{Following Proposition 3 and Lemma 1 in \cite{Zhu2010},  for a connected graph that each user has at least one neighbor, given a finite number of consensus iterations, e.g., $I_c$, all users are able to reach consensus on the global message. In other words, all SCMA users can achieve the same diversity order as that of the centralized process \cite{Yuan2018}, i.e.,} 
\begin{align}
     \mathbf{m}_{\mathbf{s},j}^{e,\text{T},I_c}[i] \rightarrow \frac{1}{J}\sum_{j=1}^J\mathbf{m}_{\mathbf{s},j}^{e,\text{T},I_c}[i], \quad
     \frac{1}{\mathbf{C}_{\mathbf{s},j}^{e,\text{T},I_c}[i,i]} \rightarrow \frac{1}{J}\sum_{j=1}^J \frac{1}{\mathbf{C}_{\mathbf{s},j}^{e,\text{T},I_c}[i,i]}.
\end{align}
Then the local extrinsic mean and covariance matrix of user $j$ are updated by $\mathbf{m}_{\mathbf{s},j}^{e,\text{T},I_c}$ and $\mathbf{C}_{\mathbf{s},j}^{e,\text{T},I_c}$, respectively, which are fed to SCMA decoder.

However, when exchanging extrinsic information between users, users' links may suffer from additive noise. To tackle this problem, a vanishing parameter $\alpha$ is introduced and (\ref{consensus_perfect}) is reformulated by \cite{Yuan2018}
\begin{equation}
\mathbf{\theta}_{i,j}^{c+1}=\mathbf{\theta}_{i,j}^c+\alpha^c\sum_{g\in\mathcal{S}_j}\gamma_{jg}\left(\mathbf{\theta}_{i,g}^c+\mathbf{\omega}_{jg}^c-\mathbf{\theta}_{i,j}^c\right),  
\label{consensus_noise}
\end{equation}
where $\mathbf{\omega}_{jg}$ is the additive noise on the link between user $j$ and user $g$.

\subsection{Reduced belief consensus-based method}
The method described in the previous subsection requires each user to broadcast all extrinsic information to neighboring users in each cross-domain iteration, which may result in significant energy consumption. To tackle this problem, a reduced energy consumption method is presented in this subsection.

The basic idea is to share small values in $\mathbf{C}_{\mathbf{s}}^{e,\text{T}}$ and their corresponding means, since larger values in $\mathbf{C}_{\mathbf{s}}^{e,\text{T}}$ indicate less accurate estimates of the L-MMSE equalizer, and such information passes between users with limited performance gains.
Specifically, for each belief consensus iteration, we diagonalize the covariance matrices due to the i.i.d. assumption, i.e. $\mathbf{c}_{\mathbf{s},j}^{e,\text{T},c} = \text{diag}\{\mathbf{C}_{\mathbf{s},j}^{e,\text{T},c}\}$. Then we sort $\mathbf{c}_{\mathbf{s},j}^{e,\text{T},c}$ in ascending order and store the corresponding indices. Let $\tilde{\mathbf{c}}_{\mathbf{s},j}^{e,\text{T},c}$ be the sorted vector and $\mathcal{I}$ be the corresponding indices set. Define the sharing rate $0 \le r_c \le 1$ that determines how much local information should be shared to nearby users. We only share the most reliable local information to nearby users. The first $\left\lfloor MN \times r_c \right\rfloor$ values in $\tilde{\mathbf{c}}_{\mathbf{s},j}^{e,\text{T},c}$ and the corresponding mean values are considered as reliable local information that is worth sharing.
Hence, the shared mean $\tilde{\mathbf{m}}_{\mathbf{s},j}^{e,\text{T},c}$ and covariance matrix $\tilde{\mathbf{C}}_{\mathbf{s},j}^{e,\text{T},c}$ are given by
\begin{align}
    \tilde{\mathbf{m}}_{\mathbf{s},j}^{e,\text{T},c} = \mathbf{m}_{\mathbf{s},j}^{e,\text{T},c}[\mathcal{I}_{[r_c]}],
    \quad 
    \tilde{\mathbf{C}}_{\mathbf{s},j}^{e,\text{T},c}[i,i] = \mathbf{c}_{\mathbf{s},j}^{e,\text{T},c}[i],
    \label{reduced}
\end{align}
where $\mathcal{I}_{[r_c]}$ denotes the first $\left\lfloor MN \times r_c \right\rfloor$ elements of the set $\mathcal{I}$ and $0 \le i \le \left\lfloor MN \times r_c \right\rfloor - 1$. {Note that the transmission of $\mathcal{I}_{[r_c]}$ can be made with negligible bandwidth increase, e.g., we can use additional decimal places to represent indices.}
According to the belief consensus recursion, $\mathbf{\theta}_{i,j}^{c}$ is given as $\mathbf{\theta}_{i,j}^{c}=\left[\tilde{\mathbf{m}}_{\mathbf{s},j}^{e,\text{T},c}[i]/\tilde{\mathbf{C}}_{\mathbf{s},j}^{e,\text{T},c}[i,i], 1/\tilde{\mathbf{C}}_{\mathbf{s},j}^{e,\text{T},c}[i,i]\right]^T$.

In this way, the extrinsic information to be shared is reduced, i.e., we only need to transmit part of the information depending on the sharing rate $r_c$, which leads to energy consumption reduction. {In practice, we may use a flag variable to indicate whether the algorithm applies the reduced belief consensus-based method or not.}

\subsection{Separate structure}
The separate structure is shown as Fig. \ref{Separated structure}. Different from the multi-layer OTFS-SCMA detector shown in Fig. \ref{Multi-layer Detector}, another scheme of distributed cooperation is to share the \textit{a posteriori} information from the DD domain after $L_\text{max}$ cross domain iterations.
Specifically, we perform belief consensus followed by an additional MPA decoding after the local OTFS-SCMA cross-domain detection. This scheme is named as \textit{separate} cross-domain and DCD rather than the \textit{joint} structure in Fig. \ref{Multi-layer Detector}, since the cooperation is performed after cross-domain detection. The cooperative process of the separated structure is expected to bring some diversity gains, but the improvement of the overall performance is limited since its cooperative process does not exploit the further gains from cross-domain message passing iteration as the joint structure does. Therefore, the performance of the separated structure should be worse than that of the joint structure under the same DCD settings.

\section{The fixed point analysis of the proposed OTFS-SCMA detector via state evolution}

In this section, we investigate the fixed point of the proposed OTFS-SCMA cross-domain detection algorithm for sufficiently large $MN$ by using \textit{state evolution} \cite{Ma2015}. {Note that the SE derived in \cite{Li2021} may not be directly applied here, due to the fact that the message passing process and NLE of the proposed detector are fundamentally different from that in \cite{Li2021}. Consequently, the SE process should be carefully re-derived.}


{Define the error terms in the $l$-th iteration of cross domain as $\mathbf{h}(l)\equiv\hat{\mathbf{s}}(l)-\mathbf{s}$ and $\mathbf{q}(l)\equiv\hat{\mathbf{x}}(l)-\mathbf{x}$. We assume that $\mathbf{h}(l)$ consists of IID entries independent of $\mathbf{H}_{\text{T}}$ and $\mathbf{n}$, and $\mathbf{q}(l)$ consists of IID zero-mean Gaussian entries independent of $\mathbf{x}$. Then we define two error measures as}
\begin{equation}
    v_{\mathbf{s}}^{p,\text{T}}(l) = \frac{1}{MN}\mathbb{E}\left\{ \|\mathbf{h}(l)\|^2 \right\}, \quad
    v_{\mathbf{x}}^{p,\text{DD}}(l) = \frac{1}{MN}\mathbb{E}\left\{ \|\mathbf{q}(l)\|^2 \right\}.
\end{equation}
With the i.i.d. assumption of both the DD domain symbols and the time domain symbols, the covariance matrices are diagonal. Thus, the measures $v_{\mathbf{s}}^{p,\text{T}}(l)$ and $v_{\mathbf{x}}^{p,\text{DD}}(l)$ can be given by
\begin{equation}
    v_{\mathbf{s}}^{p,\text{T}}(l)=\lim_{MN\rightarrow \infty}{\frac{1}{MN}\text{Tr}\left(\mathbf{C}_{\mathbf{s}}^{p,\text{T}}\right)}, \quad     
    v_{\mathbf{x}}^{p,\text{DD}}(l)=\lim_{MN\rightarrow \infty}{\frac{1}{MN}\text{Tr}\left(\mathbf{C}_{\mathbf{x}}^{p,\text{DD}}\right)}.
\end{equation}
Given the \textit{a priori} measures $v_{\mathbf{s}}^{a,\text{T}}(l)$, defined as $v_{\mathbf{s}}^{a,\text{T}}(l) \equiv \lim_{MN\rightarrow \infty}{\frac{1}{MN}\text{Tr}\left(\mathbf{C}_{\mathbf{s}}^{a,\text{T}}\right)}$, for the $l$-th iteration,
the SE for OTFS-SCMA cross-domain-based detection is defined by the following recursion
:

\textit{L-MMSE equalizer:}
\begin{equation}
    \begin{aligned}
     v_{\mathbf{s}}^{p,\text{T}}(l) = v_\mathbf{s}^{a,\text{T}}(l) -  
        \frac{\left(v_\mathbf{s}^{a,\text{T}}(l)\right)^2}{MN}\text{Tr}\left(\mathbf{H}_\text{T}^\text{H}\left(v_\mathbf{s}^{a,\text{T}}(l)\mathbf{H}_\text{T}\mathbf{H}_\text{T}^\text{H}+N_0\mathbf{I}_{MN}\right)^{-1} \mathbf{H}_\text{T}\right),
    \end{aligned}
\end{equation}

\textit{From the time domain to the DD domain:}
\begin{equation}
    v_{\mathbf{x}}^{a,\text{DD}}(l) = v_{\mathbf{s}}^{a,\text{DD}}(l)=v_{\mathbf{s}}^{e,\text{T}}(l)=\left(\frac{1}{v_{\mathbf{s}}^{p,\text{T}}(l)}- \frac{1}{v_{\mathbf{s}}^{a,\text{T}}(l)}  \right)^{-1},
\end{equation}
where $v_{\mathbf{x}}^{a,\text{DD}}(l) = v_{\mathbf{s}}^{a,\text{DD}}(l)$ is satisfied due to the unitary transformation in (\ref{ext_T2DD_m}), and the extrinsic measure (state) $v_{\mathbf{s}}^{e,\text{T}}(l)$ can be obtained by some manipulations from (\ref{ext_T_C}).

\textit{SCMA MPA decoder:}
\begin{align}
    v_{\mathbf{x},j}^{p,\text{DD}}(l)
     =\mathbb{E}\left\{ \left[X_{i,j}[k] - \mathbb{E}\left\{ X_{i,j}[k] + \sqrt{v_{\mathbf{x}}^{a,\text{DD}}(l)}Z \right\}\right]^2 \right\} 
     = \lim_{MN\rightarrow\infty}\frac{1}{MN}\text{Tr}(\mathbf{C}_j^p),\quad \forall j,
    \label{v_x_j}
\end{align}
where $v_{\mathbf{x},j}^{p,\text{DD}}(l)$\footnote{ Since the MPA decoder is a near-optimal method to approximate the solution of the MAP detection, the error measure of MPA decoder can be given in the MMSE form of (\ref{v_x_j}) under the Gaussian assumption.} denotes the error measure of MPA decoder for each user $j$, $Z$ is the AWGN sample with $Z\sim N(0,1)$ and is independent of $X_{i,j}[k]$, and the expectation in (\ref{v_x_j}) is with respect to the index $i$ and $k$ with $0\le i \le \frac{MN}{K}-1$ and $1\leq k\leq K$. 
{Unfortunately, a close form of $v_{\mathbf{x},j}^{p,\text{DD}}(l)$ may be infeasible. But $v_{\mathbf{x},j}^{p,\text{DD}}(l)$ can be regarded as a function of SNR with the SCMA system over the AWGN channels, i.e., $v_{\mathbf{x},j}^{p,\text{DD}}(l) = f(SNR)$, and this function can be obtained by Monte Carlo simulation.}

\textit{Superimposed codewords reconstruction:}
\begin{equation}
    v_{\mathbf{x}}^{p,\text{DD}}(l)=\lim_{MN\rightarrow \infty}{\frac{1}{MN}\text{Tr}\left(\mathbf{C}_{\mathbf{x}}^{p,\text{DD}}\right)}\overset{\underset{\text{(a)}}{}}{=}\left(\sum_{j=1}^j \frac{1}{v_{\mathbf{x},j}^{p,\text{DD}}(l)} \right)^{-1},
    \label{v_p_DD}
\end{equation}
where (a) is obtained by some simple manipulations from (\ref{C_x_p}).

\textit{From the DD domain to the time domain:}
\begin{equation}
    v_{\mathbf{s}}^{p,\text{DD}}(l) = v_{\mathbf{x}}^{p,\text{DD}}(l),
\end{equation}
\begin{equation}
    v_{\mathbf{s}}^{a,\text{T}}(l+1)=v_{\mathbf{s}}^{e,\text{DD}}(l)=\left(\frac{1}{v_{\mathbf{s}}^{p,\text{DD}}(l)}- \frac{1}{v_{\mathbf{s}}^{a,\text{DD}}(l)}  \right)^{-1},
\end{equation}
where $v_{\mathbf{s}}^{p,\text{DD}}(l)=v_{\mathbf{x}}^{p,\text{DD}}(l)$ is satisfied due to the unitary transformation in (\ref{p_DD2T_C}).

Based on the above analysis, the state evolution from state $v_{\mathbf{s}}^{a,\text{T}}(l)$ to $v_{\mathbf{s}}^{a,\text{T}}(l+1)$ is now well-defined. Also, the MSE for the proposed OTFS-SCMA detector is predicted as 
\begin{equation}
    \text{MSE}(l)=\frac{1}{MN}\mathbb{E}\left\{ \|\mathbf{q}(l)\|^2 \right\}=v_{\mathbf{x}}^{p,\text{DD}}(l),
    \label{MSE}
\end{equation}
since our goal is to detect the SCMA codewords in the DD domain.

We next derive the fixed point of the state evolution.

\textbf{Property 1}: When the algorithm is converged, the average of \textit{a posteriori} variance with respect to the time domain estimates and the DD domain detection outputs share the same value, i.e.,
\begin{equation}
v_\mathbf{x}^{p,\text{DD}}=v_\mathbf{s}^{p,\text{T}}.
\end{equation}
\textit{Proof:} If the algorithm is converged, the values of the states $v_\mathbf{s}^{a,\text{T}}(l)$ and $v_\mathbf{s}^{a,\text{T}}(l+1)$ will not change with the increase of the iteration number. Therefore, when the algorithm is converged, $v_{\mathbf{s}}^{a,\text{T}}(l+1)=v_{\mathbf{s}}^{a,\text{T}}(l)$ is satisfied, yielding
\begin{equation}
v_{\mathbf{s}}^{a,\text{T}}(l+1) 
= \left(\frac{1}{v_{\mathbf{s}}^{p,\text{DD}}(l)}- \frac{1}{v_{\mathbf{s}}^{a,\text{DD}}(l)}  
\right)^{-1} = \left(\frac{1}{v_{\mathbf{s}}^{p,\text{DD}}(l)} - \frac{1}{v_{\mathbf{s}}^{p,\text{T}}(l)} + \frac{1}{v_{\mathbf{s}}^{a,\text{T}}(l)} \right)^{-1},
\end{equation}
\begin{equation}
    \frac{1}{v_{\mathbf{s}}^{a,\text{T}}(l+1)} - \frac{1}{v_{\mathbf{s}}^{a,\text{T}}(l)} = \frac{1}{v_{\mathbf{s}}^{p,\text{DD}}(l)}-\frac{1}{v_{\mathbf{s}}^{p,\text{T}}(l)},
\end{equation}
\begin{equation}
    \quad v_{\mathbf{s}}^{p,\text{DD}}(l) = v_{\mathbf{s}}^{p,\text{T}}(l).
\end{equation}
This completes the proof of Property 1. \hfill $\blacksquare$

Property 1 illustrates that when the proposed algorithm converges, both time domain OTFS symbol detection and DD domain SCMA codeword decoding can provide the same accuracy regarding the data recovery. Furthermore, since the proposed cross-domain OTFS-SCMA detector follows the principle of error orthogonality, the proposed detector can converge to the Bayes optimality if there is exactly one fixed point for SE of it \cite{Liu2022}. 
{In other words, the proposed method has the potential of achieving Bayes optimality. This is, however, difficult for the two-stage detection in \cite{Deka2021} since the LMMSE estimator may not be optimal for the superimposed SCMA codewords which generally follow a complex distribution. Furthermore, when the channel is not well equalized, due to the error propagation, it could degrade the decoding performance of the subsequent MPA part in the two-stage detector, especially in the presence of fractional Doppler. By contrast, the proposed method improves the detection performance via iterations between the equalizer and the decoder with the aid of the error orthogonality principle. }

\section{Simulation results}
In this section, we carry out numerical simulations to validate the BER rate performance and convergence of the proposed OTFS-SCMA cross-domain detector.

To save the space, we summarize the simulation parameters in Table \ref{sim_parameters}\footnote{The codebook in \cite{Huang2021} has the largest MED (equals $1.3$) among the known SCMA codebooks that have been proposed, to the best of our knowledge. The Huawei codebook \cite{Nikopour2013} has large MPD and performs well over Rayleigh channels in OFDM-SCMA systems. The insight presented in Subsection III-E can be verified by comparing these two codebooks.}. {The channel suffers from fractional Doppler.} We assume the channel state information (CSI) is perfectly known at the receiver.

We first consider different OTFS-SCMA detectors, including the single-layer OTFS-SCMA cross-domain-based detector described in Fig. \ref{Detector} (termed as ``Cross domain''), {the multi-layer OTFS-SCMA detector described in Fig. \ref{Multi-layer Detector} (termed as ``Sche. 1'')}, the separate cross-domain and distributed cooperative detector described in Section IV-C (termed as ``Sche. 2'') and the two-stage detector proposed in \cite{Deka2021}. The number of cross-domain iterations $L_{\text{max}}$, SCMA MPA iterations $I_q$, belief consensus iterations $I_c$ in Sche. 1 and belief consensus iterations $I_c$ in Sche. 2 are respectively set to be $5$, $10$, $2$ and $10$, which leads to the same total number of belief consensus iterations of Sche. 1 and Sche. 2. 
{Without loss of generality, the neighboring sets $\mathcal{S}_j,\forall j,$ are fixed in our simulations, which are given by $
    \mathcal{S}_1 = \{6,2,3\}, \ \mathcal{S}_2 = \{1,3,5\}, \ \mathcal{S}_3 = \{1,2,4\}, \ 
    \mathcal{S}_4 = \{3,5,6\}, \ \mathcal{S}_5 = \{2,4,6\}, \ \text{and}\  \mathcal{S}_6 = \{1,4,5\},
 $
forming a connected graph.}

\begin{table*}[t]
    \caption{Simulation parameters}
    \label{sim_parameters}
    \resizebox{\textwidth}{25mm}{
    \begin{tabular}{l|l|l|l}
    \hline
    \textbf{Parameter}    & \textbf{Value}                                                                       & \textbf{Parameter}                                                                                & \textbf{Value}                                                                                                                     \\ \hline
    Bandwidth             & $B=10$ MHz                                                                               & \begin{tabular}[c]{@{}l@{}}Maximum Doppler index \\ (with fractional shifts)\end{tabular} & $k_{\nu_\text{max}}=6$                                                                                                                          \\ \hline
    Frame duration        & $T_f=1$ ms                                                                                 & Maximum delay index                                                                       & $l_{\tau_{\text{max}}}=3$                                                                                                                          \\ \hline
    Delay spread          & 1.6 us                                                                               & SCMA setting                                                                              & \begin{tabular}[c]{@{}l@{}}$K\times J=4\times 6$, $M_{\text{mod}}=4$\\ with factor graph Fig. \ref{Factor graph}   \end{tabular} \\ \hline
    Doppler spread        & 8 KHz                                                                                & SCMA Codebooks                                                                                & \cite{Huang2021} and \cite{Nikopour2013}                                                                                     \\ \hline
    Number of subcarriers & $M=16$                                                                                   & Number of cross-domain iterations                                                         & $L_\text{max}=5$                                                                                                                          \\ \hline
    Number of time slots  & $N=8$                                                                                    & Number of MPA decoding iterations                                                         & $I_q=10$                                                                                                                         \\ \hline
    Number of paths       & $P=4$                                                                                    & Number of belief consensus iterations                                                     & \begin{tabular}[c]{@{}l@{}}$I_c=2$ for Sche.1 \\ $I_c=10$ for Sche.2   \end{tabular}                                                                                           \\ \hline
    Channel               & Generated by (\ref{H_T}) and (\ref{H_DD})                                            & Monte Carlo (stop when $5000$ bits encountered)                                                                               & 20000 frames                                                                                                               \\ \hline
    \end{tabular}}
    \vspace{-0.5cm}
\end{table*}

\begin{figure}[t]
    \begin{minipage}[t]{0.48\linewidth}
      \centering
      \includegraphics[width=2.8in]{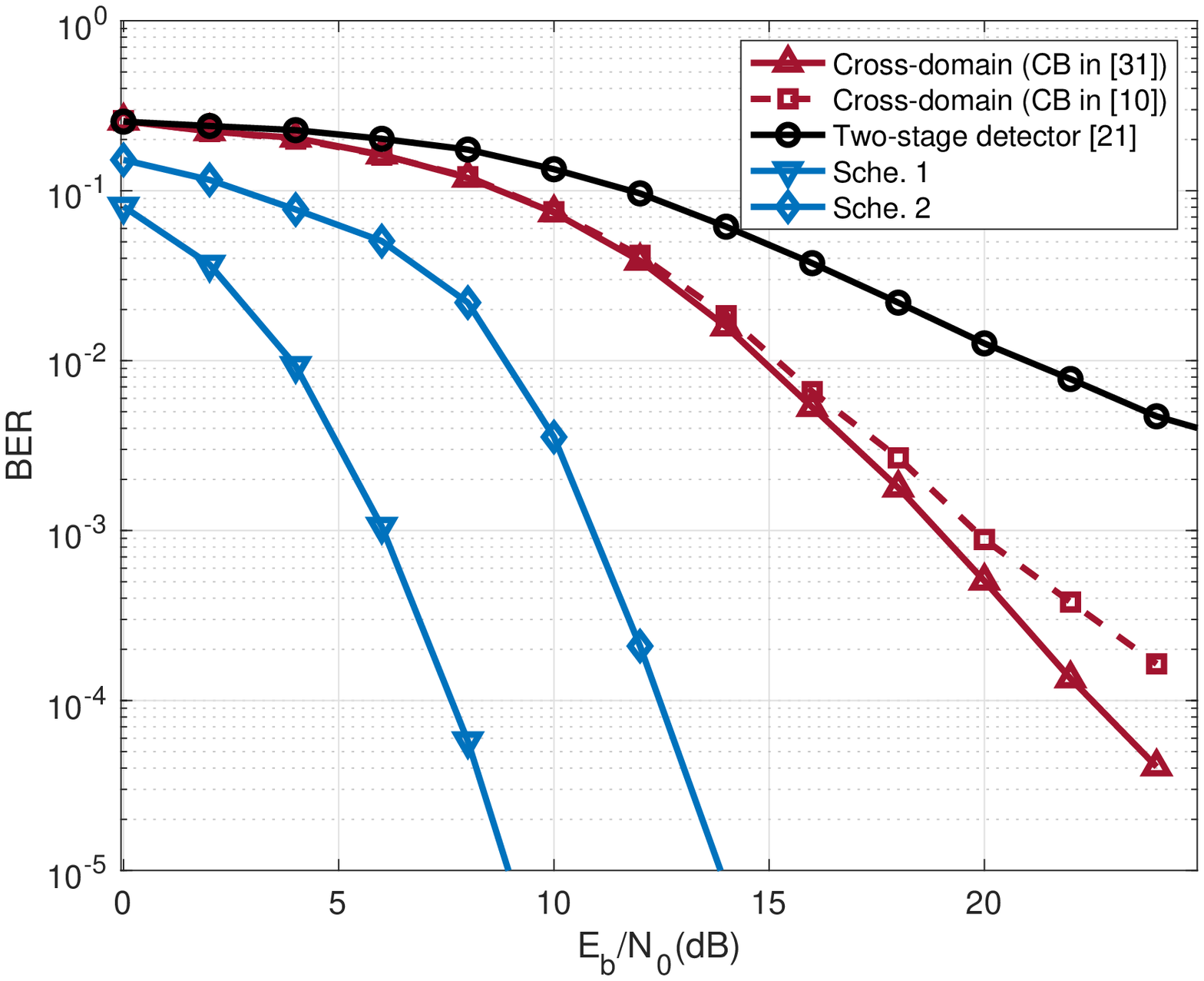}
      \caption{The BER performance of OTFS-SCMA systems with different detectors. The ``CB'' here denotes the ``codebook''. For simulations without codebook notation, the codebook of \cite{Huang2021} is employed by default.}
      \label{BER_comparison}
    \end{minipage} \quad
    \begin{minipage}[t]{0.45\linewidth}
      \centering
      \includegraphics[width=2.8in]{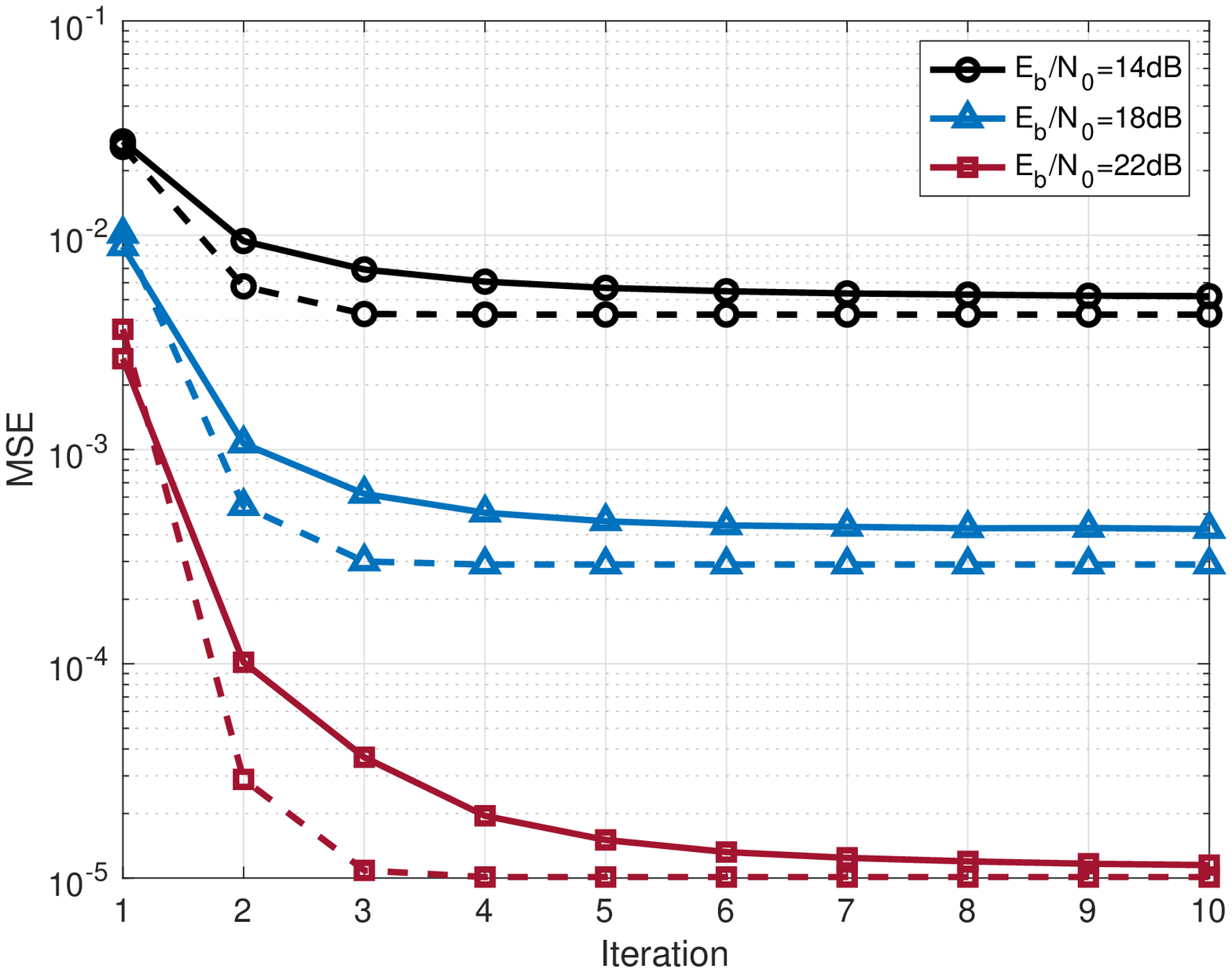}
      \caption{Simulated (solid line) and predicted (dashed line) MSEs in the DD domain.}
      \label{MSE_convergence}
    \end{minipage}
    \vspace{-7mm}
\end{figure}

Fig. \ref{BER_comparison} compares the uncoded BER performance with the above-mentioned detectors with two different SCMA codebooks. We can observe that the two-stage detector proposed in \cite{Deka2021} suffers from poor BER performance, since the fractional Doppler shifts make the channel equalization worse, which brings serious performance degradation to the subsequent SCMA decoding. This is the drawback of the two-stage detector to separately perform equalization and decoding in the DD domain without any iterations. 
The proposed OTFS-SCMA cross-domain detector (the blue line and green line) avoids the above drawback and outperforms the two-stage detector in \cite{Deka2021} significantly, thanks to the near ML detection in the DD domain and the cross-domain iterations. 
Moreover, under the proposed cross-domain detector, the codebooks of \cite{Huang2021} with large MED and small MPD outperforms the codebook of \cite{Nikopour2013} that enjoys large MPD but small MED, in high $E_b/N_0$ regions. 
This verifies the insight described in Subsection \uppercase\expandafter{\romannumeral3} E. 
Thus, the codebooks of \cite{Huang2021} are employed by default due to its enhanced error performance.
Next, we focus on the results of Sche. 1 and Sche. 2. The simple introduction of distributed cooperation after cross domain detection, i.e. Sche.2 (the red line), brings in dramatic performance gains. With the aid of cross-domain message passing and the multi-layer structure, Sche. 1 (the yellow line) can further improve the performance even though the total number of belief consensus iteration is the same as Sche. 2, leading to about $5$dB gain at $\text{BER}=10^{-4}$ compared to Sche. 2. Both Sche. 1 and Sche. 2 achieve substantial diversity gains from other downlink users and thus achieve significant performance gains compared to the other two schemes.

{Fig. \ref{MSE_convergence} compares simulated MSEs with SE prediction for the single-layer OTFS-SCMA detector in the DD domain, where the MSE is defined in (\ref{MSE}).  First, we can observe that the simulated MSEs coincide with the predicted MSEs by SE, which demonstrates that the performance of the proposed algorithm can be characterized by SE. Furthermore, the MSEs first decrease with the increasing number of iterations and then saturates close to $10^{-5}$ when $E_b/N_0=22$ dB, indicating that the proposed OTFS-SCMA detector can indeed converge after a few cross-domain iterations (less than 5 iterations).}


\begin{figure}
    \begin{minipage}[t]{0.32\linewidth}
      \centering
      \includegraphics[width=2.1in]{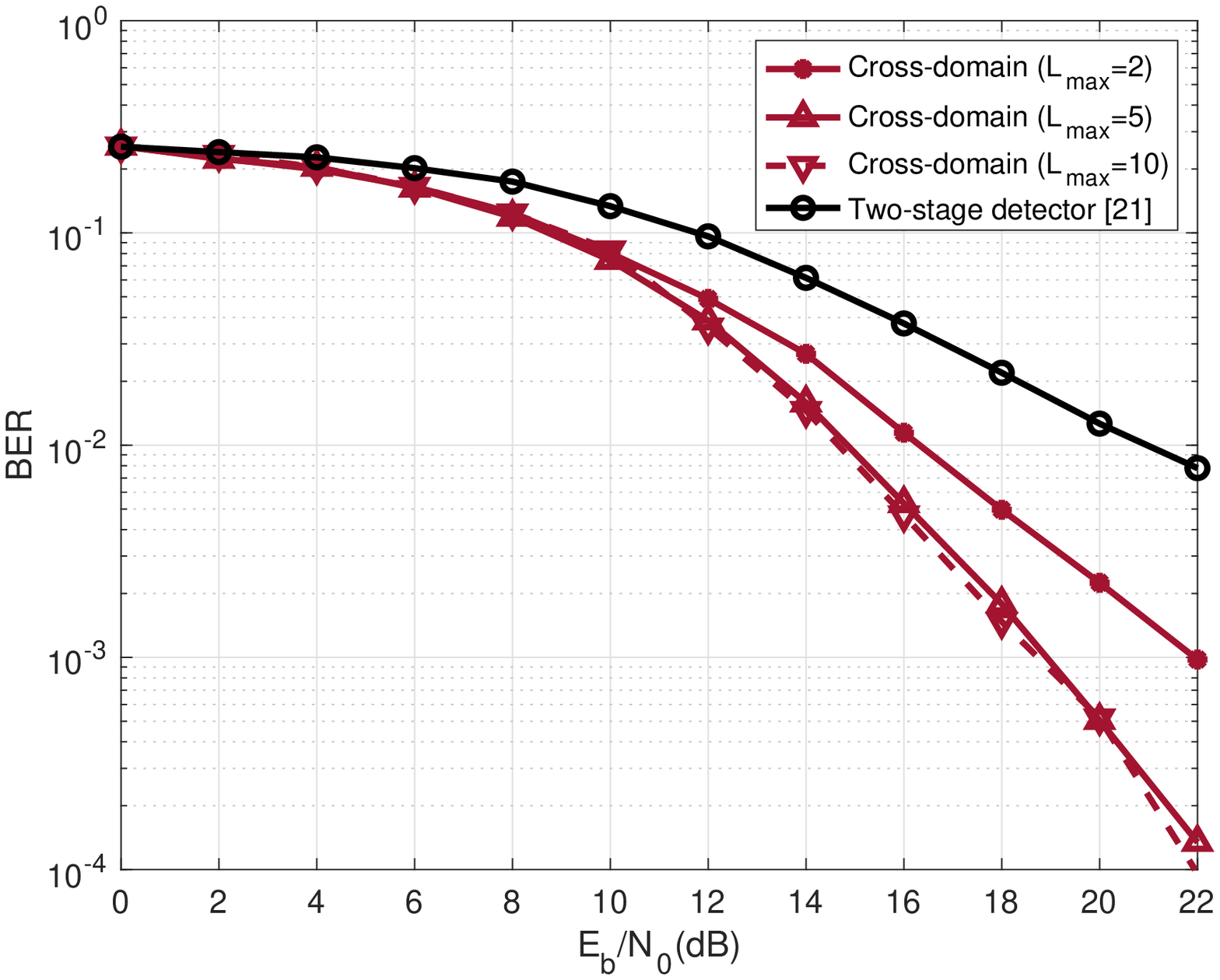}
      \caption{BER comparison of the proposed single-layer OTFS-SCMA detector with \cite{Deka2021}.}
      \label{Different_iter_BER}
    \end{minipage}  \ 
    \begin{minipage}[t]{0.32\linewidth}
      \centering
      \includegraphics[width=2.1in]{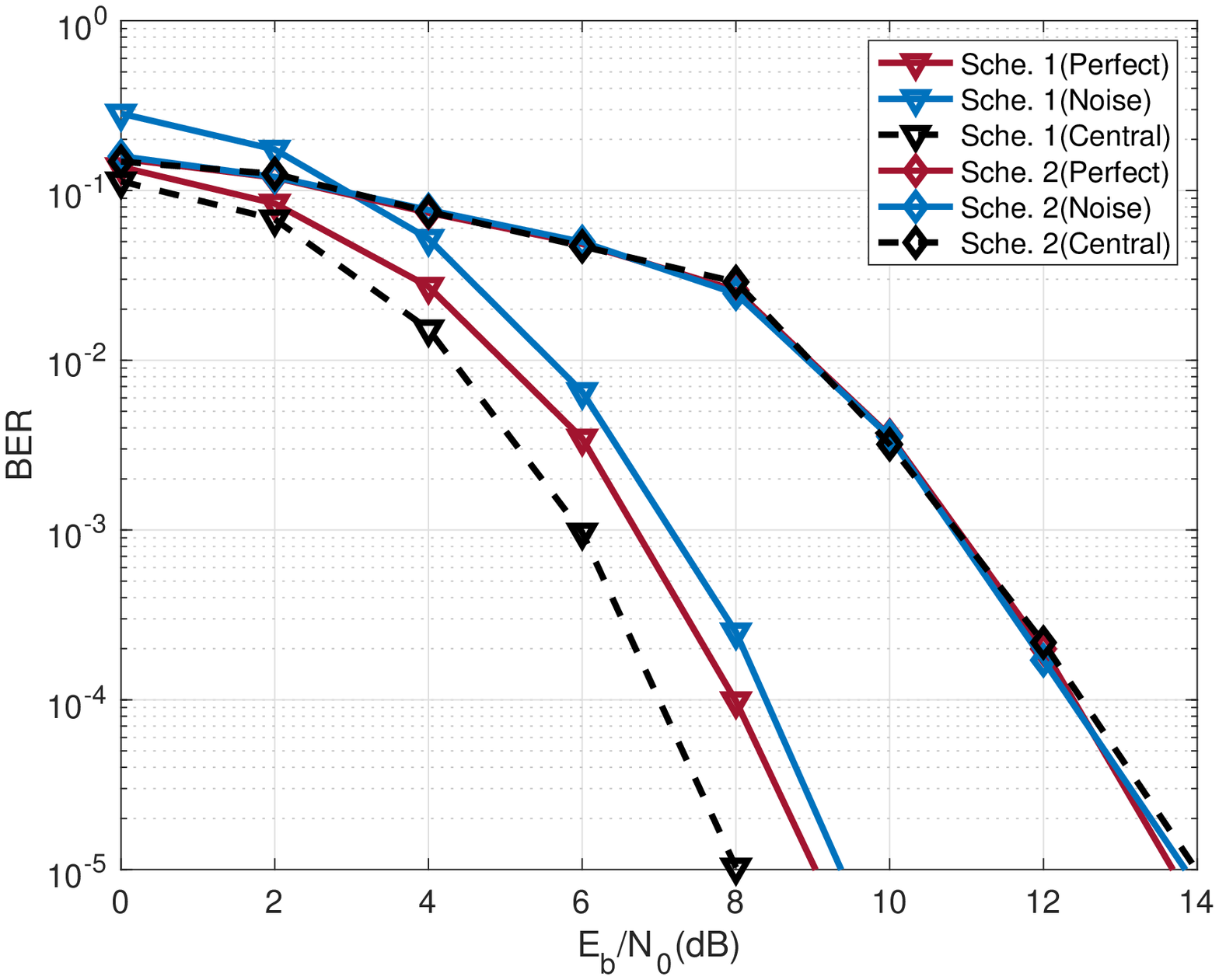}
      \caption{BER comparison of the multi-layer OTFS-SCMA detector and  the separate cross-domain and distributed cooperative detector with central, perfect and noisy inter-user links.}
      \label{BER_distributed_cooperation}
    \end{minipage} \ 
    \begin{minipage}[t]{0.32\linewidth}
        \includegraphics[width=2.1in]{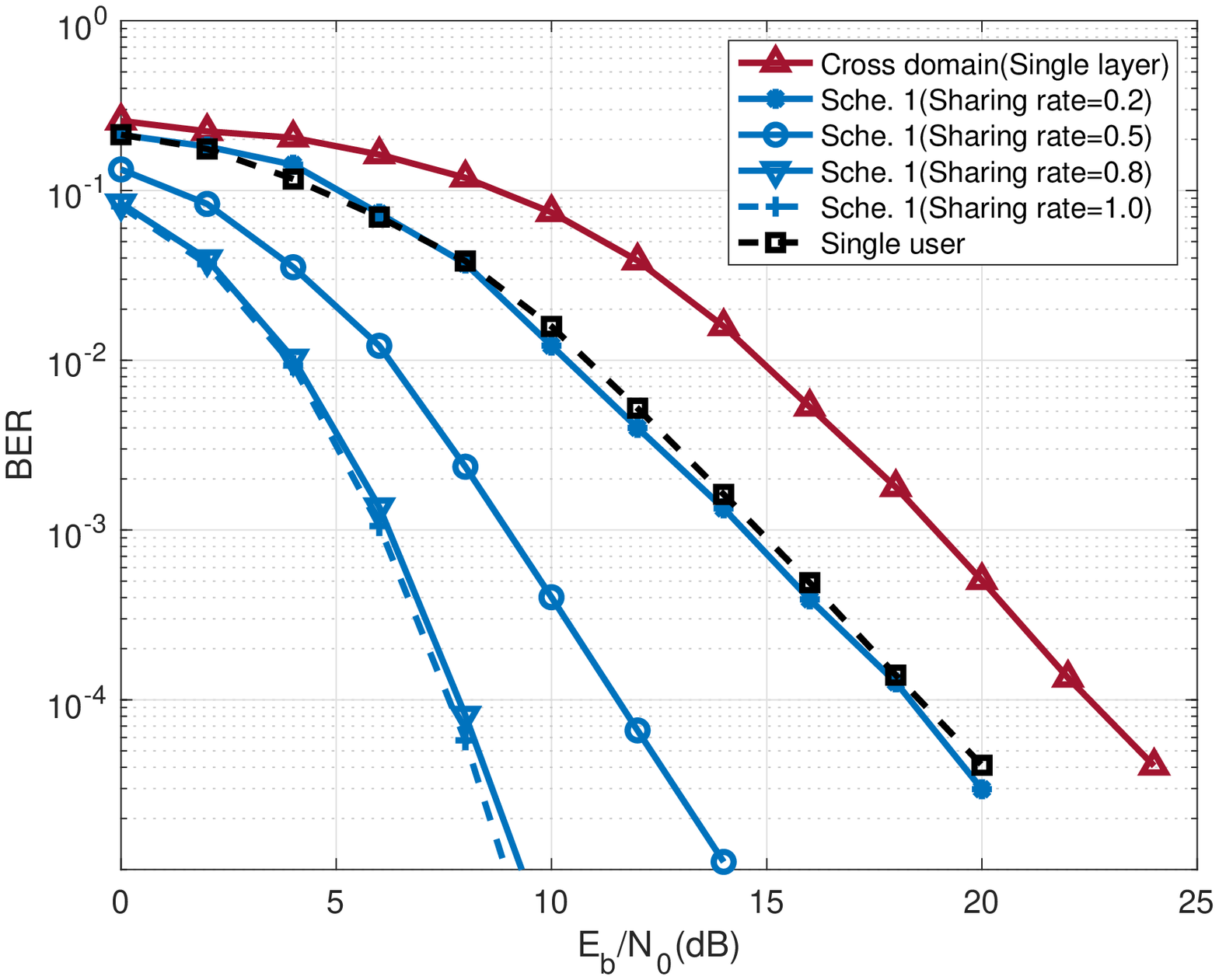} 
        \caption{BER comparison of the multi-layer OTFS-SCMA detector with different sharing rates. The dotted line represents the single-user cross-domain OTFS system \cite{Li2021} that transmits QPSK symbols.}
        \label{BER_reduced_scheme}
    \end{minipage}
    \vspace{-7mm}
\end{figure}



Fig. \ref{Different_iter_BER} shows the BER performance of the single-layer OTFS-SCMA detector with different number of cross-domain iterations and that of \cite{Deka2021}. The BER results in Fig. \ref{Different_iter_BER} consistent with the MSE performance shown in Fig. \ref{MSE_convergence}. It can be observed that in Fig. \ref{Different_iter_BER}, very few
cross-domain iterations, e.g. $L_{\text{max}}=2$, can lead to significant BER performance gains compared to that of \cite{Deka2021}.


To illustrate the robustness of the proposed distributed cooperation scheme, we consider different schemes of inter-user links, including central, perfect and noisy inter-user links. Specifically, the central link means that the local information of user $j$ is broadcast to all the other users, and the perfect link means that the noise term in (\ref{consensus_noise}) is zero whereas the noise term in (\ref{consensus_noise}) in the noisy link is a random value drawn from the Gaussian distribution with zero mean and variance of one. 
Fig. \ref{BER_distributed_cooperation} shows the BER performance of two distributed cooperation schemes (Sche. 1 and Sche. 2) with different inter-user links. Although the Sche. 2 performs almost the same in different inter-user link conditions, the Sche. 1 outperforms Sche. 2 in all conditions under the same total number of belief consensus iterations ($I_c^{\text{total}}=10$) when $E_b/N_0 \ge 4$dB. Due to fewer number of belief consensus iterations in each cross-domain iteration for Sche. 1 in our simulations, e.g., $I_c=2$, the noise in (\ref{consensus_noise}) can not be vanished in low $E_b/N_0$ regions. For the same reason, the Sche. 1 can only achieve full user diversity gains in the central link, while Sche. 2 can achieve full user diversity gains in both the central link and distributed links (i.e., the perfect link and the noisy link). Despite Sche. 1 cannot achieve full user diversity gains in distributed links, its error performance is remarkably impressive, with less than $1$ dB difference from the central scheme.


Fig. \ref{BER_reduced_scheme} shows the BER comparisons of the multi-layer OTFS-SCMA detector with different sharing rates in perfect links. The single-layer OTFS-SCMA detector and the single-user cross-domain OTFS system proposed in \cite{Li2021} that transmits 4QAM symbols are also considered. One can observe that when the sharing rate equals $0.8$, the performance is close to the scheme of sharing all local information. When only half ($r=0.5$) local information is shared to nearby users, it suffers from about $5$ dB performance degradation. That said, it has significantly outperformed the single-user OTFS cross-domain system with about $7$ dB gain at $\text{BER}=10^{-4}$. The most impressive result is that when the sharing rate equals $0.2$, meaning that only $20\%$ local information is shared to nearby users, the error performance of Sche. 1 is close to that of the single-user cross-domain system and even better than that in high $E_b/N_0$ regions. Even though $r=0.2$, the Sche. 1 still achieve about $5$ dB gain compared with the single-layer detector at $\text{BER}=10^{-4}$. This observation demonstrates that the proposed multi-layer OTFS-SCMA cross-domain detector can support massive multiple access, leading to an increased spectral efficiency (with the overloading factor that is larger than $1$) and can outperform the single-user cross-domain system at a factional energy and delay cost. In a real-world environment, we can adjust the sharing rate according to the BER requirements.

\section{Conclusions}
In this paper, we have proposed a novel downlink code-domain NOMA scheme by integrating OTFS and SCMA. 
We first proposed a single-layer cross-domain detection algorithm for OTFS-SCMA systems.
In the cross-domain OTFS-SCMA systems, we have shown that SCMA codebooks designed for AWGN channels with large MEDs are desired.
Then, with the introduction of the cooperative network, we have developed a joint cross-domain and DCD algorithm (the multi-layer OTFS-SCMA detector) to achieve large user diversity gains. 
Furthermore, the fixed point of the proposed OTFS-SCMA detection algorithm has been analyzed based on the state evolution technique.
Our numerical results have shown that the proposed schemes can converge with significant performance gains compared with previous methods.
Moreover, the simulation results demonstrate that the proposed cooperative schemes can outperform single-user systems and support massive multiple access communications in high-mobility environments with excellent error performance at very small energy cost.

Despite the great performance of the proposed downlink OTFS-SCMA systems, it is interesting to investigate the uplink OTFS-NOMA which is challenging for rapid and accurate estimation of the CSI for different users.

\bibliographystyle{IEEEtran}

\end{document}